\documentclass[nofootinbib]{revtex4}
\usepackage{graphicx}
\usepackage{dcolumn}
\usepackage{amsmath,amssymb,epsfig}
\usepackage{paralist}
\usepackage{comment}
\usepackage{graphicx}
\usepackage{wrapfig}
\usepackage{multirow}
\usepackage{color,soul}
\usepackage{suffix}
\usepackage{mathtools}

\allowdisplaybreaks

\renewcommand{\vec}[1]{\boldsymbol{\mathrm{#1}}}
\newcommand\ddfrac[2]{{\displaystyle\frac{\displaystyle #1}{\displaystyle #2}}}

\begin{document}

\title{Imaging faint sources with the extended solar gravitational lens}

\author{Slava G. Turyshev$^{1}$, Viktor T. Toth$^2$}

\affiliation{\vskip 3pt$^1$Jet Propulsion Laboratory, California Institute of Technology,\\
4800 Oak Grove Drive, Pasadena, CA 91109-0899, USA}

\affiliation{\vskip 3pt $^2$Ottawa, Ontario K1N 9H5, Canada}

\date{\today}

\begin{abstract}

We consider resolved imaging of faint sources with the solar gravitational lens (SGL) while treating the Sun as an extended gravitating body. We use our new diffraction integral that describes how a spherical electromagnetic wave is modified by the static gravitational field of an extended body, represented by series of multipole moments characterizing its interior mass distribution. Dominated by the solar quadrupole moment, these deviations from spherical symmetry significantly perturb the image that is projected by the Sun into its focal region, especially at solar equatorial latitudes. To study the optical properties of the quadrupole SGL, we develop an approximate solution for the point spread function of such an extended lens. We also derive semi-analytical expressions to estimate signal levels from extended targets. With these tools, we study the impact of solar oblateness on imaging with the SGL. Given the small value of the solar quadrupole moment, the majority of the signal photons arriving from an extended target still appear within the image area projected by the monopole lens. However, these photons are scrambled, thus reducing the achievable signal-to-noise ratio during image recovery process (i.e., deconvolution). We also evaluate the spectral sensitivity for high-resolution remote sensing of exoplanets with the extended SGL. We assess the impact on image quality and demonstrate that despite the adverse effects of the quadrupole moment, the SGL remains uniquely capable of delivering high-resolution imaging and spectroscopy of faint, small and distant targets, notably terrestrial exoplanets within $\sim$30--100 parsec from us.

\end{abstract}


\maketitle

\section{Introduction}
\label{sec:intro}

With their impressive capabilities, current and upcoming fleets of space-based optical telescopes will accelerate the discoveries of terrestrial exoplanets and will also be able to detect signs of habitability\footnote{\url{https://exoplanets.nasa.gov/}}. This information is expected from transit spectroscopy and reflected imaging observations that will probe the atmospheric composition of these newly discovered worlds. However, visiting these targets is out of question: there are no extant or foreseeable advances in propulsion, power and communication technologies to allow for interstellar travel. Remote sensing is our only viable option. To confirm the presence of higher-order life on the surface of a distant exoplanet, ideally we would have access to direct multi-pixel imaging and high-resolution spectroscopy \cite{Seager2014}.  Such imaging of terrestrial exoplanets, using conventional astronomical instruments (i.e.,  telescopes and interferometers) is a challenging endeavor \cite{Lagrange:2014,Hinkley:2021}. These targets are small, they are at large distances from us, they are not self-luminous and their light arrives on a strong noise background due to their host star and exozodiacal light \cite{Traub-Oppenheimer:2010,Gaudi:2013}.

To overcome these challenges, we can try to increase the sensitivity and resolution of our optical instruments. This can be achieved by developing ever larger telescopes, preferably placing them outside the Earth's atmosphere. Over the last quarter century we have successfully operated the Hubble Space Telescope\footnote{\url{https://en.wikipedia.org/wiki/Hubble_Space_Telescope}} (HST) with its 2.4~m primary mirror, working at optical wavelengths. Its long-anticipated successor, the James Webb Space Telescope\footnote{\url{https://en.wikipedia.org/wiki/James_Webb_Space_Telescope}} (JWST) began operation in 2022, collecting images and spectroscopic data with its 6.5-m primary mirror, working in the near-infrared (near-IR) band. This observatory is expected to bring us exciting transit spectroscopy data on many exoplanets in our stellar neighborhood. Even larger space telescopes with $\sim$15--25~m apertures may be built in the 2030s \cite{Astro2020}. Still, not even these instruments will allow us to directly image exoplanetary surfaces or conduct spatially resolved surface spectroscopy, which would be needed to unambiguously confirm higher-order life and habitation. The main obstacle is the diffraction limit: in order to see an Earth-like planet as just one resolved pixel from a distance of $z_0=100$ light-years (ly), a diffraction-limited telescope aperture or interferometric baseline of $1.22 \lambda (z_0/2R_\oplus)\sim90.5(\lambda/1\,\mu{\rm m})$ km would be needed.  Resolved multipixel images with $n$ linear pixels require proportionally larger operating configurations.

Clearly, such instrument systems are not feasible anytime soon, if ever. Although the new generation of science instruments will allow us to gather more data from observation of unresolved objects as they transit over their host stars, these instruments will not offer any details about the surfaces of these planets. We will not be able to see their continents and oceans, study their weather, understand the planetary topography, or see signs of technological activity. We will learn nothing about the possible presence of a civilization with these instruments. To discover, confirm, and study advanced life on a distant world, a radically new approach is needed.

One possible solution that is within reach at our current level of technology is the solar gravitational lens (SGL) \cite{Turyshev-Toth:2017,Turyshev-Toth:2019-extend,Turyshev-Toth:2020-photom,Turyshev-Toth:2020-image,Turyshev-Toth:2022-broad-SNR,Turyshev-Toth:2022-mono-SNR}. According to the general theory of relativity \cite{Einstein-1911,Einstein-1916,Einstein:1936}, the gravitational field of a massive body, manifesting in the form of spacetime curvature, affects light propagation in the vicinity of that body \cite{Singe-1960,Liebes:1964,Schneider-Ehlers-Falco:1992,Refsdal-Surdej:1994,Wambsganss:1998}. The curvature affects an electromagnetic (EM) wave by induces a phase delay that depends on the impact parameter. As a result, the EM wavefront is deflected, focused behind the body, leading to significant gain or light amplification \cite{Turyshev-Toth:2017}. Any massive body acts as a lens. The larger the mass, the stronger is the lens.

The parameter determining the strength of a gravitational lens is its Schwarzschild radius $r_g=2GM/c^2$, where $M$ is the mass of the body. Among solar system bodies, the Sun is the heaviest and most dynamically stable object, providing a lens with unique optical properties. For the Sun $r_g\simeq 2.95$~km, which provides the SGL with its impressive optical properties, including its peak light amplification, $\mu_0=4\pi^2 r_g/\lambda=1.17 \times 10^{11} (1~\mu{\rm m}/\lambda)$ and angular resolution, $\delta\theta\simeq 0.38 (\lambda/\sqrt{2r_g\overline{z}})=0.10(\lambda/1~\mu{\rm m})(650\,{\rm AU}/\overline{z})^\frac{1}{2}$ nanoarcsecond \cite{Turyshev-Toth:2017,Turyshev-Toth:2020-extend}.

The focal region of the SGL begins at the heliocentric distance of $\overline z = R_\odot^2/2r_g=547.8 (b/R_\odot)^2$ astronomical units (AU), where $R_\odot$ is the solar radius \cite{vonEshleman:1979,Turyshev-Toth:2017}. The success of the Voyager 1 spacecraft, as it continues to operate at distances beyond 158~AU after 45 years in deep space\footnote{\url{https://voyager.jpl.nasa.gov/mission/status/}}, demonstrates that the SGL's focal region is accessible with current technologies \cite{Helvajian2022}. Placing at the focal region of the SGL a 1-m telescope that is equipped with a coronagraph to block light from our Sun and treating the solar corona as the main source of noise, yields a strong signal-to-noise ratio (SNR) for imaging of faint objects. Technical challenges notwithstanding, the SGL appears to be our only realistic means for direct high-resolution imaging and spectroscopy of terrestrial exoplanets in our stellar neighborhood within $\sim$100--300 light-years ($\sim$30--100 parsec).

Recognizing the unique value of the SGL for the quest of finding, confirming and remotely investigating life outside our home planet, we studied its optical properties extensively \cite{Turyshev-Toth:2017,Turyshev-Toth:2019-extend,Turyshev-Toth:2020-photom,Turyshev-Toth:2020-image,Turyshev-Toth:2021-all-regions}. We developed a wave-optical treatment \cite{Born-Wolf:1999} of the SGL and modeled light propagation in the vicinity of the Sun \cite{Turyshev-Toth:2019,Turyshev-Toth:2019-plasma}. We addressed the impact of the solar corona, treating it as the main source of stochastic noise \cite{Turyshev-Toth:2020-extend,Toth-Turyshev:2020}.  This work allowed us to validate the imaging capabilities of the SGL and develop a mission concept that would be able to deliver resolved imaging and spectroscopic data within a realistic mission timeframe \cite{Turyshev-Toth:2022-mono-SNR,Turyshev-Toth:2022-broad-SNR}.

The Sun is not a perfect gravitational monopole. We extended our model to treat the Sun as a body with a gravitational field that is characterized by an infinite set of gravitational multipole moments \cite{Turyshev-Toth:2021-multipoles,Turyshev-Toth:2021-all-regions}. 
We showed how each of these multipole moments affects the point-spread function (PSF) of the SGL, leading to the formation of caustics in the image plane \cite{Turyshev-Toth:2021-caustics,Turyshev-Toth:2021-imaging,Turyshev-Toth:2021-quartic}. We recognized that the although the Sun is an extended body, its nonspherical deformations are very small and are dominated by the quadrupole moment $J_2$. Given the small value of this moment, light received from an extended body in an image plane in the focal region of the SGL, still falls mostly within the image area defined by the monopole lens, but it is scrambled, which affects image recovery (deconvolution), amplifying noise, reducing the SNR of the recovered image. Understanding the impact of the solar quadrupole on the quality of image recovery is essential, which is the primary motivation for this paper.

The paper is organized as follows:
Section \ref{sec:im-form} summarizes the image formation process with the extended SGL. Section \ref{sec:quad-PSF} discusses the quadrupole PSF. We develop an approximate solution for the diffraction integral, dealing separately with the region inside and outside the astroid caustic boundary formed by the quadrupole moment. In Section \ref{sec:intense-in}, we develop analytical expressions describing the intensity distribution and power received at the focal plane of the imaging telescope. In Section \ref{sec:signals}, we evaluate the SGL-amplified signals and estimate the spectral SNR for imaging and spectroscopy of exoplanets with the extended SGL. In Section \ref{sec:conc}, we summarize results and discuss possible next steps in the investigation of the SGL.
To keep the main text streamlined, we placed some detailed calculations in the Appendices.

\section{Wave optical treatment of extended gravitational lenses}
\label{sec:im-form}

We consider the propagation of a monochromatic EM wave that originates at a target of radius of $R_\oplus$ positioned at the distance of $z_0$ from the Sun (we rely on results presented in \cite{Turyshev-Toth:2020-extend,Turyshev-Toth:2021-multipoles,Turyshev-Toth:2021-imaging}). The wave travels in the direction toward the Sun, where it is focused and amplified by the SGL. To capture the image formed by the EM waves from the target, we position an imaging telescope in the SGL focal region, at the heliocentric distance $\overline z$ and near the primary optical axis (the line that connects the center of the target with that of the Sun and extends into the focal region of the SGL, see \cite{Turyshev-Toth:2017,Turyshev-Toth:2019-extend,Turyshev-Toth:2020-photom,Turyshev-Toth:2020-image,Turyshev-Toth:2020-extend}) in what is known as the region of strong interference of the SGL. Next, we introduce two-dimensional coordinates to describe points in the source plane, $\vec x'$; the position of the telescope in the image plane,  $\vec x_0$; points in the image plane within the telescope's aperture,  $\vec x$; and points in the optical telescope's focal plane ${\vec x}_i$. These are given as  follows:
\begin{eqnarray}
\{{\vec x}'\}&\equiv& (x',y')=\rho'\big(\cos\phi',\sin\phi'\big)=\rho'{\vec n}',
\label{eq:x'}\\
\{{\vec x}_0\}&\equiv& (x_0,y_0)=\rho_0\big(\cos\phi_0,\sin\phi_0\big)=\rho_0{\vec n}_0, \label{eq:x0}\\
\{{\vec x}\}&\equiv& (x,y)=\rho\big(\cos\phi,\sin\phi\big)=\rho\,{\vec n}, \label{eq:x}\\
 \{{\vec x}_i\}&\equiv& (x_i,y_i)=\rho_i\big(\cos\phi_i,\sin\phi_i\big)=\rho_i{\vec n}_i.
  \label{eq:p}
\end{eqnarray}

We rely on  (\ref{eq:x'})--(\ref{eq:p}), but slightly redefining them  by introducing ${\vec x}_0=-({\overline z}/{z_0}){\vec x}_0'$, which allows us to use
{}
\begin{eqnarray}
{\vec x}''={\vec x}'-{\vec x}'_0\equiv \rho''{\vec n}''= \rho'' \big(\cos\phi'',\sin\phi''\big).
  \label{eq:coord2}
\end{eqnarray}

To image faint, distant objects with the SGL, we represent an imaging telescope by a  convex thin lens with aperture $d$ and focal distance $f$. We position the telescope at a point with coordinates ${\vec x}_0$ in the image plane in the strong interference region of the lens (see discussion in \cite{Turyshev-Toth:2019-extend,Turyshev-Toth:2020-extend}).  To stay within the image, ${\vec x}_0$ is within the range:  $|{\vec x}_0|+d/2\leq r_\oplus$, where $r_\oplus=(\bar z/z_0)R_\oplus$ is the radius of the image.

We introduce the following notations for the two spatial frequencies $\alpha$ and $\eta_i$, and a useful scale ratio $\beta$:
{}
\begin{eqnarray}
\alpha&=&k \sqrt{\frac{2r_g}{\overline z}}, \qquad \eta_i=k\frac{\rho_i}{f}, \qquad \beta=\frac{\overline z}{{z}_0},
  \label{eq:alpha-mu}\\
 \vec \alpha&=&(\alpha_x,\alpha_y)=\alpha \big(\cos\phi_\xi,\sin\phi_\xi\big)=\alpha \vec n_\xi, \qquad \vec \eta_i=\eta_i \vec n_i,
 \label{eq:vec_alpha-eta}
\end{eqnarray}
where $k=2\pi/\lambda$ is the wavenumber of an EM wave and $\vec n_\xi$ is the unit vector in the direction of the light ray's vector impact parameter \cite{Turyshev-Toth:2017,Turyshev-Toth:2020-extend,Turyshev-Toth:2021-multipoles}.

With these definitions, the intensity distribution at the focal plane of the imaging telescope is given by
  {}
\begin{eqnarray}
I_{\tt }({\vec x}_i,{\vec x}_0) &=&
\frac{1}{z^2_0}\mu_0 \Big(\frac{kd^2}{8f}\Big)^2\hskip-4pt
\iint d^2{\vec x}''  B_{\tt s}({\vec x}'')    {\cal A}^2({\vec x}_i,{\vec x}''), \qquad \mu_0=2\pi kr_g,
  \label{eq:pow-blur}
\end{eqnarray}
where $B_{\tt s}({\vec x}'') $ is the source's surface brightness and ${\cal A}({\vec x}_i,{\vec x}'')$ is the Fourier-transformed complex amplitude of the EM wave at the focal plane of the imaging telescope, which is given as
{}
\begin{eqnarray}
{\cal A}({\vec x}_i,{\vec x}'')&=&
 \frac{1}{2\pi}\int_0^{2\pi} d\phi_\xi \Big(\frac{2J_1(\alpha d\, \hat u(\phi_\xi, \vec x_i))}{\alpha d\, \hat u(\phi_\xi,\vec x_i)}\Big)\times\nonumber\\
 &&\times\, \exp\Big[-ik\Big(\sqrt{\frac{2r_g}{\bar z}} \beta \rho''\cos(\phi_\xi-\phi'') +
2r_g\sum_{n=2}^\infty \frac{J_n}{n} \Big(\frac{R_\odot }{\sqrt{2r_g\bar z}}\Big)^n\sin^n\beta_s\cos[n(\phi_\xi-\phi_s)]\Big)\Big],
  \label{eq:amp-blur3*}
\end{eqnarray}
where the $J_n$ in (\ref{eq:amp-blur3*}) are the dimensionless zonal harmonic coefficients characterizing the mass distribution within the Sun,  while $(\beta_s,\phi_s)$ are the target's position in a heliocentric spherical coordinate system that is aligned with the solar axis of rotation (see \cite{Turyshev-Toth:2021-multipoles,Turyshev-Toth:2021-caustics,Turyshev-Toth:2021-quartic,Turyshev-Toth:2021-imaging,Turyshev-Toth:2021-all-regions} for details.) In addition, $J_1(x)$ is the Bessel function of the first kind and $u(\phi_\xi,\vec x_i)$ is the normalized spatial frequency, which, with the definitions (\ref{eq:alpha-mu})--(\ref{eq:vec_alpha-eta}), has the structure
{}
\begin{eqnarray}
\hat u(\phi_\xi,\vec x_i)=
\big| \vec \alpha + \vec \eta_i\big|/2\alpha
&=&
\Big\{{\textstyle\frac{1}{4}}\Big(1-\frac{\eta_i}{\alpha}\Big)^2+\frac{\eta_i}{\alpha}\cos^2[{\textstyle\frac{1}{2}}\big(\phi_\xi-\phi_i\big)] \Big\}^\frac{1}{2}.
  \label{eq:upm}
\end{eqnarray}

From (\ref{eq:upm}), we can see that at the exact location of the Einstein ring that forms in the telescope's focal plane, i.e., when $\eta_i =\alpha$ or $\rho_i=f\sqrt{{2r_g}/{\overline z}}$, the spatial frequency $\hat u(\phi_\xi,\vec x_i)$ collapses to $\hat u_{\tt ER}(\phi_\xi,\vec x_i)=\cos[{\textstyle\frac{1}{2}}\big(\phi_\xi-\phi_i\big)]$. That means that for any position in the focal plane with azimuthal coordinate $\phi_i$ along the Einstein ring, there will be a particular angle $\phi_\xi$ that will result in $\hat u_{\tt ER}(\phi_\xi,\vec x_i)=0$. It is only at those points along the Einstein ring, the ratio $2J_1(x)/x$ in (\ref{eq:amp-blur3*}) reaches it largest value of $2J_1(x)/x\rightarrow 1$, making the largest contribution to the overall integral (\ref{eq:amp-blur3*}).

The integral (\ref{eq:pow-blur}) must be evaluated for two different regions corresponding to the telescope pointing within the image and outside of it, as was done in \cite{Turyshev-Toth:2020-photom}. The principal technical challenge is the evaluation of the integral (\ref{eq:amp-blur3*}) that represents a Fourier-transform of the EM field amplitude. In \cite{Turyshev-Toth:2022-broad-SNR}, we addressed that challenge in the case of a monopole lens (i.e., when\footnote{The dipole moment, $J_1$, vanishes when the origin of the coordinate system coincides with the Sun's center-of-mass.} $J_{n\geq 2}=0$), opening the path for analytical or semi-analytical treatments. We now use a similar approach in the case of a weakly aspherical axisymmetric lens with a small quadrupole mass moment (i.e., when $J_2\ne 0, J_{n>2}=0$).

\section{The quadrupole PSF of the extended SGL}
\label{sec:quad-PSF}

The Sun is an axisymmetric rotating body with north-south symmetry. Its mass distribution, therefore, can be fully represented using even multipole moments $J_{2n}$ contributing to the integral (\ref{eq:amp-blur3*}). The solar multipole moments are well-determined using  interplanetary spacecraft tracking data, yeilding $J_2=(2.25\pm0.09)\times 10^{-7}$ \cite{Park-etal:2017},  and $J_4=-4.44\times 10^{-9}$, $J_6=-2.79\times 10^{-10}$, $J_8=1.48\times 10^{-11}$ \cite{Roxburgh:2001}. As was shown in \cite{Turyshev-Toth:2021-multipoles,Turyshev-Toth:2021-caustics,Turyshev-Toth:2021-quartic}, for the SGL the contribution from the solar $J_2$ dominates, and when studying the qualitative properties of the SGL PSF, contributions from higher moments may be safely neglected. Therefore, we now consider only the quadrupole SGL, formally characterized by $0<J_2\ll 1, J_{n>2}=0$.

\subsection{Fourier-transformed amplitude of the EM field}

In the case of a quadrupole SGL, the Fourier-transformed amplitude ${\cal A}({\vec x}_i,{\vec x}'')$ of the EM field given by (\ref{eq:amp-blur3*}) takes the form \cite{Turyshev-Toth:2021-multipoles,Turyshev-Toth:2021-caustics,Turyshev-Toth:2021-quartic}:
{}
\begin{eqnarray}
{\cal A}({\vec x}_i,{\vec x}'')&=&
 \frac{1}{2\pi}\int_0^{2\pi} d\phi_\xi \Big(\frac{2J_1(\alpha d\, \hat u(\phi_\xi, \vec x_i))}{\alpha d\, \hat u(\phi_\xi,\vec x_i)}\Big)
 \exp\Big[-i\Big(\alpha \beta \rho''\cos(\phi_\xi-\phi'') +
\beta_2\cos[2(\phi_\xi-\phi_s)]\Big)\Big],
  \label{eq:amp-quad}
\end{eqnarray}
where $\alpha$ and $\beta$ are from (\ref{eq:alpha-mu}) and $u(\phi_\xi,\vec x_i)$ is given by (\ref{eq:upm}). The values of $\alpha$ and $\beta_2$ are estimated as
{}
\begin{eqnarray}
\alpha&=&k\sqrt\frac{2r_g}{\overline z}=48.97\, \Big(\frac{1\,\mu{\rm m} }{\lambda}\Big)\Big(\frac{650{\rm AU} }{\overline z}\Big)^\frac{1}{2}~{\rm m}^{-1},
\label{eq:zerJ}\\
\beta_2 &=&kr_gJ_2 \Big(\frac{R_\odot }{\sqrt{2r_g \overline z}}\Big)^2\sin^2\beta_s=3518.34\,\sin^2\beta_s\, \Big(\frac{J_2}{2.25\times 10^{-7}}\Big)\Big(\frac{1\,\mu{\rm m} }{\lambda}\Big)\Big(\frac{650{\rm AU} }{\overline z}\Big)~{\rm rad}.
\label{eq:beta2}
\end{eqnarray}

Despite its nice, elegant, and compact form, the integral (\ref{eq:amp-quad}) is not known to have an analytic solution.  (Even for the monopole case we had to treat this integral using the method of stationary phase \cite{Turyshev-Toth:2022-mono-SNR}.) This is why, in \cite{Turyshev-Toth:2021-multipoles,Turyshev-Toth:2021-imaging,Turyshev-Toth:2021-all-regions} we evaluated it numerically, and in \cite{Turyshev-Toth:2021-caustics} we studied a semi-analytical form of the quadrupole SGL using the algebraic solution of a quartic equation in \cite{Turyshev-Toth:2021-quartic}.   All of these approaches (especially the approach based on the quartic equation) can be used to develop either numerical or semi-analytical estimates of the SNR. However, numerical results in particular provide less insight for instrument and mission development. Thus, another treatment of the integral (\ref{eq:amp-quad}) is needed.

Similarly to the approach taken in \cite{Turyshev-Toth:2022-mono-SNR}, we may evaluate the integral (\ref{eq:amp-quad}) using the method of stationary phase.  With the rapidly varying phase given as
{}
\begin{eqnarray}
\varphi(\phi_\xi)=-\Big(\alpha\beta \rho''\cos(\phi_\xi-\phi'')+
\beta_2\cos[2(\phi_\xi-\phi_s)]\Big),
  \label{eq:ph-quad}
\end{eqnarray}
for that, we compute the first and second derivatives of this expression:
{}
\begin{eqnarray}
\varphi'(\phi_\xi)&=&\alpha\beta \rho''\sin(\phi_\xi-\phi'')+2
\beta_2\sin[2(\phi_\xi-\phi_s)],
  \label{eq:ph-quad-der1}\\
\varphi''(\phi_\xi)&=&\alpha\beta \rho''\cos(\phi_\xi-\phi'')+4\beta_2\cos[2(\phi_\xi-\phi_s)].
  \label{eq:ph-quad-der2}
\end{eqnarray}

The phase is stationary when $\varphi'(\phi_\xi)=0$. Solving this equation algebraically using Cardano's quartic solution we obtain four roots for $\phi_\xi^{[n]}, n\in[1,4]$ \cite{Turyshev-Toth:2021-quartic}. This quartic-based solution may be used to evaluate the integral (\ref{eq:amp-quad}) and then use the result in (\ref{eq:pow-blur}), as was done in \cite{Turyshev-Toth:2021-quartic}. However, though valid, the solution still has to be evaluated numerically. As an alternative, we now develop another analytic solution to (\ref{eq:amp-quad}) that may provide much needed insight into the overall imaging properties of the extended SGL and the achievable SNR.

As was shown in \cite{Turyshev-Toth:2021-caustics} the integral (\ref{eq:amp-quad}) rapidly oscillates and has a sharp transition boundary in the form of the astroid caustic \cite{Berry-Upstill:1982,Berry:1992} that, using polar coordinates $(\beta\rho'',\phi'')$ in the image plane, is given by
{}
\begin{eqnarray}
\beta\rho''_{\tt ac}(\phi)&=&\frac{4\beta_2}{\alpha}\Big(\sin^\frac{2}{3}(\phi''-\phi_s)+\cos^\frac{2}{3}(\phi''-\phi_s)\Big)^{-\frac{3}{2}}.
  \label{eq:caust}
\end{eqnarray}

Using (\ref{eq:zerJ})--(\ref{eq:beta2}), we estimate the unscaled magnitude of the astroid caustic in the image plane given by (\ref{eq:caust}) as
{}
\begin{eqnarray}
\frac{4\beta_2}{\alpha}&=&2J_2 
\frac{R^2_\odot }{\sqrt{2r_g\bar z}}
\sin^2\beta_s=287.39\, \sin^2\beta_s\Big(\frac{650~{\rm AU}}{\overline z}\Big)^\frac{1}{2}~~{\rm m}.
  \label{eq:4beta2}
\end{eqnarray}
Therefore, for a given target, the size of the quadrupole caustic of the SGL is determined by the angle $\beta_s$ (corresponding to the sky position of the source), and the distance of the image plane from the Sun. When the observed target is in the direction to the solar axis of rotation, i.e., $\beta_s\simeq 0$ or $\beta_s\simeq \pi$, the caustic collapses, resulting in an approximately monopole pattern. Otherwise, we must account for its presence in the image plane, especially when the magnitude of the caustic is much lager then the telescope aperture, i.e., when $d\ll 4\beta_2/\alpha$. The numerical values of the parameters $\alpha$ and $\beta_2$  (as well as their ratio ${4\beta_2}/{\alpha}$) characterize the challenge of evaluating the integral (\ref{eq:amp-quad}): any change in the radial position in the image plane, $\rho''$, results in rapid variations of the phase of the integrand.

Refraction of light in the solar corona may  reduce the size of the astroid caustic.  As was shown in \cite{Turyshev-Toth:2019,Turyshev-Toth:2019-plasma}, due to its negative refractive index, the solar corona counteracts he gravitational deflection of light by bending the light trajectories outwards and effectively pushing the focal area of the SGL to larger heliocentric distances. Oblateness of the steady-state solar corona may yield a contribution to the astroid caustic that counteracts the action of the solar quadrupole moment, thus effectively reducing the size of the astroid caustic (\ref{eq:4beta2}). Although, as was seen in \cite{Turyshev-Toth:2019,Turyshev-Toth:2019-plasma},  such an effect on the phase delay of a light ray at optical wavelengths may be rather small,  it may be prove to be useful and, thus, it needs to be fully evaluated. The relevant work is ongoing and the results will be reported.

The caustic (\ref{eq:caust}) separates two regions with markedly different behavior for the integral (\ref{eq:amp-quad}). In the region inside the caustic the quadrupole term (i.e., $\propto\beta_2$) dominates. The region outside the caustic is dominated by the monopole term (i.e., $\propto \alpha\beta\rho''$). This observation allows us to develop approximate solutions for each of these two regions. Based on the caustic properties studied in \cite{Turyshev-Toth:2021-caustics,Turyshev-Toth:2021-quartic}, the two regions are characterized by the following conditions on the parameters present in the phase of of the integral (\ref{eq:amp-quad}): $\alpha\beta\rho''<4\beta_2$ vs. $\alpha\beta\rho''\geq 4\beta_2$.  These conditions lead to iteratively developed stationary phase solutions for both of these regions.

Formally, this can be done by first computing the Poynting vector of the EM wave in the two regions with respect the caustic boundary and then computing the PSF by averaging the result, as was done, for instance, in \cite{Turyshev-Toth:2020-extend}. As the phases of the waves in these regions are very different, the mixed term that would appear in the PSF is very small and may be neglected. This is why we can characterize these two regions by two different expressions for the PSF, with each obtained from (\ref{eq:amp-quad}), evaluated in the corresponding region.

\subsection{The quadrupole-dominated region: $\alpha\beta\rho''<4\beta_2$}

\subsubsection{The stationary phase solutions}

To develop an iterative stationary phase solution to (\ref{eq:ph-quad-der1}) in the quadrupole-dominated region (i.e., within the caustic boundary $\rho''\leq \rho_{\tt ac}$), we introduce a small parameter:
{}
\begin{eqnarray}
0\leq x&=& \frac{\alpha\beta\rho''}{4\beta_2}<1.
  \label{eq:quad-small-par}
\end{eqnarray}
Clearly, when $\rho''$ takes its largest value at the caustic boundary given by (\ref{eq:caust}), this inequality is still valid. Using this parameter $x$, we may solve (\ref{eq:ph-quad-der1}) up to the terms of ${\cal O}(x^5)$, by searching for the solution for $\phi_\xi$ in the form
{}
\begin{eqnarray}
\phi_\xi&=&\phi_\xi^{[0]}+\phi_\xi^{[1]}+\phi_\xi^{[2]}+
\phi_\xi^{[3]}+\phi_\xi^{[4]}+
{\cal O}\big(x^5\big),
  \label{eq:beta-phi_xi}
\end{eqnarray}
with the zeroth order equation after substitution of (\ref{eq:beta-phi_xi}) in (\ref{eq:ph-quad-der1}), taking the form
{}
\begin{eqnarray}
\sin[2(\phi^{[0]}_\xi-\phi_s)]&=&0,
  \label{eq:phi_xi0}
\end{eqnarray}
that yields the following four solutions for $\phi^{[0]}_\xi$
{}
\begin{eqnarray}
\phi^{[0]}_\xi-\phi_s=\Big\{0,{\textstyle\frac{\pi}{2}},\pi,{\textstyle\frac{3\pi}{2}}\Big\},
  \label{eq:ph-quad-sol}
\end{eqnarray}
consistent with those found in \cite{Turyshev-Toth:2021-caustics}  at the origin of the coordinate system, i.e., where $\rho''=0$. Continuing with the iterative approach, we found four solutions of (\ref{eq:ph-quad-der1}):
 {}
\begin{eqnarray}
\phi_{\xi0}&=&
\phi_s+\frac{\alpha\beta\rho''}{4\beta_2}\sin(\phi''-\phi_s)-{\textstyle\frac{1}{2}}\Big(\frac{\alpha\beta\rho''}{4\beta_2}\Big)^2\sin[2(\phi''-\phi_s)]+\nonumber\\
&&+\,
{\textstyle\frac{1}{24}}\Big(\frac{\alpha\beta\rho''}{4\beta_2}\Big)^3\Big(9\sin[\phi''-\phi_s]+5\sin[3(\phi''-\phi_s)]\Big)-{\textstyle\frac{1}{2}}\Big(\frac{\alpha\beta\rho''}{4\beta_2}\Big)^4\sin[2(\phi''-\phi_s)]+{\cal O}\Big(\Big(\frac{\alpha\beta\rho''}{4\beta_2}\Big)^5\Big),
  \label{eq:quad-sol0}\\[4pt]
\phi_{\xi{\textstyle\frac{\pi}{2}}}&=&
{\textstyle\frac{\pi}{2}}+ \phi_s+\frac{\alpha\beta\rho''}{4\beta_2}\cos(\phi''-\phi_s)+{\textstyle\frac{1}{2}}\Big(\frac{\alpha\beta\rho''}{4\beta_2}\Big)^2\sin[2(\phi''-\phi_s)]+\nonumber\\
&&+\,
{\textstyle\frac{1}{24}}\Big(\frac{\alpha\beta\rho''}{4\beta_2}\Big)^3\Big(9\cos[\phi''-\phi_s]-5\cos[3(\phi''-\phi_s)]\Big)+{\textstyle\frac{1}{2}}\Big(\frac{\alpha\beta\rho''}{4\beta_2}\Big)^4\sin[2(\phi''-\phi_s)]+{\cal O}\Big(\Big(\frac{\alpha\beta\rho''}{4\beta_2}\Big)^5\Big),
  \label{eq:quad-solpi2}\\[4pt]
\phi_{\xi\pi}&=&
\pi+ \phi_s-\frac{\alpha\beta\rho''}{4\beta_2}\sin(\phi''-\phi_s)-{\textstyle\frac{1}{2}}\Big(\frac{\alpha\beta\rho''}{4\beta_2}\Big)^2\sin[2(\phi''-\phi_s)]-\nonumber\\
&&-\,
{\textstyle\frac{1}{12}}\Big(\frac{\alpha\beta\rho''}{4\beta_2}\Big)^3\sin[\phi''-\phi_s]\Big(7 +5\cos[2(\phi''-\phi_s)]\Big)-{\textstyle\frac{1}{2}}\Big(\frac{\alpha\beta\rho''}{4\beta_2}\Big)^4\sin[2(\phi''-\phi_s)]+{\cal O}\Big(\Big(\frac{\alpha\beta\rho''}{4\beta_2}\Big)^5\Big),
  \label{eq:quad-solpi} \\[4pt]
\phi_{\xi{\textstyle\frac{3\pi}{2}}}&=&
{\textstyle\frac{3\pi}{2}} +\phi_s-\frac{\alpha\beta\rho''}{4\beta_2}\cos(\phi''-\phi_s)+{\textstyle\frac{1}{2}}\Big(\frac{\alpha\beta\rho''}{4\beta_2}\Big)^2\sin[2(\phi''-\phi_s)]-\nonumber\\
&&-\,
{\textstyle\frac{1}{12}}\Big(\frac{\alpha\beta\rho''}{4\beta_2}\Big)^3\cos[\phi''-\phi_s]\Big(7-5\cos[2(\phi''-\phi_s)]\Big)+{\textstyle\frac{1}{2}}\Big(\frac{\alpha\beta\rho''}{4\beta_2}\Big)^4\sin[2(\phi''-\phi_s)]+{\cal O}\Big(\Big(\frac{\alpha\beta\rho''}{4\beta_2}\Big)^5\Big).
  \label{eq:quad-sol3pi2}
\end{eqnarray}

With these solutions for the stationary phase at hand, we compute the second derivatives of the phase (\ref{eq:ph-quad-der2}). The corresponding results are obtained in the form that, for compactness, we present pairwise, for $\{0,\pi\}$ and $\{{\textstyle\frac{\pi}{2}},{\textstyle\frac{3\pi}{2}}\}$:
 {}
\begin{eqnarray}
\varphi''_{\xi\{0|\pi\}}&=&4\beta_2\Big\{1\pm
\frac{\alpha\beta\rho''}{4\beta_2}\cos(\phi''-\phi_s)-\Big(\frac{\alpha\beta\rho''}{4\beta_2}\Big)^2\sin^2(\phi''-\phi_s)\pm\nonumber\\
&&\hskip 20pt \pm\,
{\textstyle\frac{5}{4}}\Big(\frac{\alpha\beta\rho''}{4\beta_2}\Big)^3\sin(\phi''-\phi_s)\sin[2(\phi''-\phi_s)]-\Big(\frac{\alpha\beta\rho''}{4\beta_2}\Big)^4\sin^2[2(\phi''-\phi_s)]+{\cal O}\Big(\Big(\frac{\alpha\beta\rho''}{4\beta_2}\Big)^5\Big)\Big\},
  \label{eq:quad-2dir0}\\[4pt]
\varphi''_{\xi\{{\textstyle\frac{\pi}{2}}|{\textstyle\frac{3\pi}{2}}\}}&=& -4\beta_2\Big\{1\mp
\frac{\alpha\beta\rho''}{4\beta_2}\sin(\phi''-\phi_s)-\Big(\frac{\alpha\beta\rho''}{4\beta_2}\Big)^2\cos^2(\phi''-\phi_s)\mp\nonumber\\
&&\hskip 20pt \mp\,
{\textstyle\frac{5}{4}}\Big(\frac{\alpha\beta\rho''}{4\beta_2}\Big)^3\cos(\phi''-\phi_s)\sin[2(\phi''-\phi_s)]-\Big(\frac{\alpha\beta\rho''}{4\beta_2}\Big)^4\sin^2[2(\phi''-\phi_s)]+{\cal O}\Big(\Big(\frac{\alpha\beta\rho''}{4\beta_2}\Big)^5\Big)\Big\},
  \label{eq:quad-2dirpi2}
  \label{eq:quad-2dir3pi2}
\end{eqnarray}
where the upper sign corresponds to the first solution in a pair.

We also compute the values for the stationary phase (\ref{eq:ph-quad}) for each of the four solutions, again presented in pairs:
 {}
\begin{eqnarray}
\varphi_{\xi\{0|\pi\}}&=&-\beta_2\Big\{1\pm
4\frac{\alpha\beta\rho''}{4\beta_2}\cos(\phi''-\phi_s)+2\Big(\frac{\alpha\beta\rho''}{4\beta_2}\Big)^2\sin^2(\phi''-\phi_s)\mp\nonumber\\
&&\hskip 20pt \mp\,
\Big(\frac{\alpha\beta\rho''}{4\beta_2}\Big)^3\sin(\phi''-\phi_s)\sin[2(\phi''-\phi_s)]+{\textstyle\frac{1}{2}}\Big(\frac{\alpha\beta\rho''}{4\beta_2}\Big)^4\sin^2[2(\phi''-\phi_s)]+{\cal O}\Big(\Big(\frac{\alpha\beta\rho''}{4\beta_2}\Big)^5\Big)\Big\},
  \label{eq:quad-2ph0}\\[4pt]
  \varphi_{\xi\{{\textstyle\frac{\pi}{2}}|{\textstyle\frac{3\pi}{2}}\}}&=& \beta_2\Big\{1\mp
4\frac{\alpha\beta\rho''}{4\beta_2}\sin(\phi''-\phi_s)+2\Big(\frac{\alpha\beta\rho''}{4\beta_2}\Big)^2\cos^2(\phi''-\phi_s)\pm\nonumber\\
&&\hskip 20pt \pm\,
\Big(\frac{\alpha\beta\rho''}{4\beta_2}\Big)^3\cos(\phi''-\phi_s)\sin[2(\phi''-\phi_s)]+{\textstyle\frac{1}{2}}\Big(\frac{\alpha\beta\rho''}{4\beta_2}\Big)^4\sin^2[2(\phi''-\phi_s)]+{\cal O}\Big(\Big(\frac{\alpha\beta\rho''}{4\beta_2}\Big)^5\Big)\Big\},
  \label{eq:quad-2phpi2}
  \label{eq:quad-2ph3pi2}
\end{eqnarray}
where again, the upper sign corresponds to the first solution in the pair.

We note that none of the expressions for the second derivative of the phase (\ref{eq:quad-2dir0})--(\ref{eq:quad-2dir3pi2}) changes signs for any values of the small parameter $x$ (given by (\ref{eq:quad-small-par})) while $x$ is within the inner region of the caustic, namely for $0\leq x<1$. These expressions stay either positive (i.e., for (\ref{eq:quad-2dir0})) or negative (i.e., (\ref{eq:quad-2dirpi2})). This fact allows us to assemble the stationary phase solution for the integral (\ref{eq:amp-quad}) within the inner caustic region.

Also note that all of the presented solutions correctly represent the behavior characteristic to the inner region of the astroid caustic: as $\rho''$ changes, the solutions move both radially and azimuthally as was observed in \cite{Turyshev-Toth:2021-multipoles,Turyshev-Toth:2021-caustics,Turyshev-Toth:2021-quartic}.

\subsubsection{The complex amplitude}

To establish the solution for the complex amplitude (\ref{eq:amp-quad}) in the inner caustic region, ${\cal A}_{\tt in}({\vec x}_i,{\vec x}'')$, we first define, for convenience, the amplitude ${\cal A}_\xi$ for each of the four solutions:
{}
\begin{eqnarray}
{\cal A}_\xi&=&
\Big(\frac{2J_1(\alpha d\, \hat u(\phi_\xi, \vec x_i))}{\alpha d\, \hat u(\phi_\xi,\vec x_i)}\Big) \qquad \rightarrow \qquad
{\cal A}_\xi=\Big\{{\cal A}_0,{\cal A}_{\frac{\pi}{2}},{\cal A}_\pi,{\cal A}_\frac{3\pi}{2}\Big\}.
  \label{eq:amp}
\end{eqnarray}
Note that each of these amplitudes is rather small, reaching their maximum value of 1 when their arguments vanish, i.e., ${\cal A}_\xi\equiv (2J_1(x_\xi)/x_\xi)\rightarrow 1$, when $x_\xi\rightarrow 0$. With the definition (\ref{eq:amp}) and the results for the phase and its second derivative obtained above, we may now present the stationary phase solution for the Fourier-transformed amplitude of the EM field given by the integral (\ref{eq:amp-quad}) in the inner part of the astroid caustic in the  following form:
{}
\begin{eqnarray}
{\cal A}_{\tt in}({\vec x}_i,{\vec x}'')&=&
 \frac{1}{\sqrt{8\pi\beta_2}}\Big\{{\cal A}_0\frac{e^{i \big(\varphi_{\xi0}+\frac{\pi}{4}\big)}}{\sqrt{\overline\varphi''_{\xi0}}}+{\cal A}_{\frac{\pi}{2}}
\frac{e^{i \big(\varphi_{\xi\pi/2}-\frac{\pi}{4}\big)}}{\sqrt{\overline\varphi''_{\xi\pi/2}}}+
{\cal A}_{\pi}\frac{e^{i \big(\varphi_{\xi\pi}+\frac{\pi}{4}\big)}}{\sqrt{\overline\varphi''_{\xi\pi}}}
+{\cal A}_{\frac{3\pi}{2}}\frac{e^{i \big(\varphi_{\xi3\pi/2}-\frac{\pi}{4}\big)}}{\sqrt{\overline\varphi''_{\xi3\pi/2}}}\Big\},
  \label{eq:amp-quad-int}
\end{eqnarray}
where $\overline\varphi''_{\xi} $ is the normalized second derivative of the phase given as $\overline\varphi''_{\xi} =\varphi''_{\xi}/4\beta_2$ with its form evident from (\ref{eq:quad-2dir0})--(\ref{eq:quad-2dir3pi2}).

To evaluate the  expression (\ref{eq:pow-blur}) in the inner part of the astroid caustic, we need to compute the square of the Fourier-transformed amplitude of the EM field, ${\cal A}_{\tt in}^2({\vec x}_i,{\vec x}'')$. We derive this expression by squaring (\ref{eq:amp-quad-int}) and using  the expressions derived for the phases (\ref{eq:quad-2ph0})--(\ref{eq:quad-2ph3pi2}) and their second derivatives (\ref{eq:quad-2dir0})--(\ref{eq:quad-2dir3pi2}). As a result, the quantity ${\cal A}_{\tt in}^2({\vec x}_i,{\vec x}'')$ may be given in the following form:
{}
\begin{eqnarray}
{\cal A}_{\tt in}^2({\vec x}_i,{\vec x}'')&=&
 \frac{1}{8\pi\beta_2} {\cal B}_{\tt in}^2({\vec x}_i,{\vec x}''),
  \label{eq:psf-quad_in-AB}
 \end{eqnarray}
where $ {\cal B}_{\tt in}^2({\vec x}_i,{\vec x}'')$ is given by a lengthy expression (\ref{eq:psf-quad_in-AB2}) that we reproduced in Appendix~\ref{app:EMamp}.

This result represents an approximation for the amplitude of the EM field in the inner part of the caustic region (see also \cite{Turyshev-Toth:2021-caustics} where we studied behavior of the  PSF of the quadratic SGL along the direction toward the cusp and that toward the fold). Compared to the monopole version of the amplitude of the EM field \cite{Turyshev-Toth:2020-extend,Toth-Turyshev:2020} which depends only on the radial coordinate $\rho''$,  expression (\ref{eq:psf-quad_in-AB2}) also depends on the azimuthal angle $\phi''$. This dependence significantly impacts the deconvolution penalty \cite{Turyshev-Toth:2020-extend,Turyshev-Toth:2022-broad-SNR}. This expression may now be used in (\ref{eq:pow-blur}) that allows us to estimate the signal intensity of the signal received at the focal plane of an imaging telescope.

\subsection{The monopole dominated region: $\alpha\beta\rho''>4\beta_2$}

\subsubsection{The stationary phase solutions}

To develop an iterative solution to the monopole dominated region, we introduce another small parameter:
{}
\begin{eqnarray}
0\leq y&=& \frac{4\beta_2}{\alpha\beta\rho''}<1.
  \label{eq:quad-small-par-y}
\end{eqnarray}
Using this parameter, we may solve (\ref{eq:ph-quad-der1}) up to the terms\footnote{As the quantity $y$ from (\ref{eq:quad-small-par-y}) diminishes rapidly as $\rho''$ increases, it is sufficient to use only the first few terms in a power series expansion.} of ${\cal O}(y^3)$, by searching for a solution for $\phi_\xi$ in the form
{}
\begin{eqnarray}
\phi_\xi&=&\phi_\xi^{[0]}+\phi_\xi^{[1]}+\phi_\xi^{[2]}+{\cal O}\big(y^3\big),
  \label{eq:beta-phi_xi-y}
\end{eqnarray}
with the zeroth order equation taking the form
{}
\begin{eqnarray}
\alpha\beta\rho''\sin\big(\phi^{[0]}_\xi-\phi''\big)&=&0,
  \label{eq:phi_xi0-y}
\end{eqnarray}
that yields the following two solutions for $\phi^{[0]}_\xi$:
{}
\begin{eqnarray}
\phi^{[0]}_\xi-\phi''=\big\{0,\pi\big\},
  \label{eq:ph-quad-sol-m}
\end{eqnarray}
consistent with those found in \cite{Turyshev-Toth:2017} for the monopole SGL. Continuing with the iterative approach, we found the following two solutions of (\ref{eq:ph-quad-der1}) that we identify as $\{\phi_{\xi0},\phi_{\xi\pi}\}$ and that are given as
 {}
\begin{eqnarray}
\phi_{\xi0}&=&
\phi''-{\textstyle\frac{1}{2}}\Big(\frac{4\beta_2}{\alpha\beta\rho''}\Big)\sin[2(\phi''-\phi_s)]+{\textstyle\frac{1}{4}}\Big(\frac{4\beta_2}{\alpha\beta\rho''}\Big)^2\sin[4(\phi''-\phi_s)]+
{\cal O}\Big(\Big(\frac{4\beta_2}{\alpha\beta\rho''}\Big)^3\Big),
  \label{eq:quad-sol0-y}\\
\phi_{\xi\pi}&=&
\pi+ \phi''+ {\textstyle\frac{1}{2}}\Big(\frac{4\beta_2}{\alpha\beta\rho''}\Big)\sin[2(\phi''-\phi_s)]+{\textstyle\frac{1}{4}}\Big(\frac{4\beta_2}{\alpha\beta\rho''}\Big)^2\sin[4(\phi''-\phi_s)]+{\cal O}\Big(\Big(\frac{4\beta_2}{\alpha\beta\rho''}\Big)^3\Big).
  \label{eq:quad-solpi-y}
  \end{eqnarray}

With these solutions for the stationary phase at hand, we compute the second derivatives of the phase (\ref{eq:ph-quad-der2}) and also compute the values for the stationary phase (\ref{eq:ph-quad}) for each of the two solutions (\ref{eq:quad-sol0-y})--(\ref{eq:quad-solpi-y}). The corresponding results are obtained in the following form:
 {}
\begin{eqnarray}
\varphi''_{\xi\{0|\pi\}}&=&\pm\alpha\beta\rho''\Big\{1\pm
\cos(\phi''-\phi_s)\Big(\frac{4\beta_2}{\alpha\beta\rho''}\Big)+{\textstyle\frac{7}{8}}\sin^2[2(\phi''-\phi_s)]\Big(\frac{4\beta_2}{\alpha\beta\rho''}\Big)^2+{\cal O}\Big(\Big(\frac{4\beta_2}{\alpha\beta\rho''}\Big)^3\Big)\Big\},
  \label{eq:quad-2dir0-y}
  \label{eq:quad-2dirpi-y}\\
\varphi_{\xi\{0|\pi\}}&=&\mp \alpha\beta\rho''\Big\{1\pm
{\textstyle\frac{1}{2}}\cos(\phi''-\phi_s)\Big(\frac{4\beta_2}{\alpha\beta\rho''}\Big)+{\textstyle\frac{3}{8}}\sin^2[2(\phi''-\phi_s)]\Big(\frac{4\beta_2}{\alpha\beta\rho''}\Big)^2+{\cal O}\Big(\Big(\frac{4\beta_2}{\alpha\beta\rho''}\Big)^3\Big)\Big\},
  \label{eq:quad-2ph0-y}
  \label{eq:quad-2phpi-y}
\end{eqnarray}
where once again, the upper sign corresponds to the first solution in a pair.

We can now use these results to assemble the stationary phase solution in the region outside the caustic.

\subsubsection{The complex amplitude}

We note that none of the expressions for the second derivative of the phase (\ref{eq:quad-2dir0-y}) change signs for any values of the small parameter $y$ (given by (\ref{eq:quad-small-par-y})) while $y$ is in the outer caustic region. These expressions stay either positive (i.e., for uppers sign in (\ref{eq:quad-2dir0-y})) or negative (i.e., for lower sign in (\ref{eq:quad-2dirpi-y})). This fact allows us to assemble the stationary phase solution for the integral (\ref{eq:amp-quad}) for the outer caustic region. For that, similar to (\ref{eq:amp}), we define the amplitude ${\cal A}_\xi$ for each of the two solutions:
{}
\begin{eqnarray}
{\cal A}_\xi&=&
\Big(\frac{2J_1(\alpha d\, \hat u(\phi_\xi, \vec x_i))}{\alpha d\, \hat u(\phi_\xi,\vec x_i)}\Big) \qquad \rightarrow \qquad
{\cal A}_\xi=\Big\{{\cal A}_0,{\cal A}_\pi\Big\},
  \label{eq:amp-y}
\end{eqnarray}
where all the quantities involved are computed with (\ref{eq:quad-sol0-y})--(\ref{eq:quad-solpi-y}).

With the definition (\ref{eq:amp-y}) and the results for the phase and its second derivative obtained above, we may now present the stationary phase solution the Fourier-transformed amplitude of the EM field given by the integral (\ref{eq:amp-quad}) outside of the astroid caustic in the  following form:
{}
\begin{eqnarray}
{\cal A}_{\tt out}({\vec x}_i,{\vec x}'')&=&
 \frac{1}{\sqrt{2\pi \alpha\beta\rho''}}\Big\{{\cal A}_0\frac{e^{i \big(\varphi_{\xi0}+\frac{\pi}{4}\big)}}{\sqrt{\overline\varphi''_{\xi0}}}+
{\cal A}_{\pi}\frac{e^{i \big(\varphi_{\xi\pi}-\frac{\pi}{4}\big)}}{\sqrt{\overline\varphi''_{\xi\pi}}}\Big\},
  \label{eq:amp-quad-int-y}
\end{eqnarray}
where $\overline\varphi''_{\xi} $ is the normalized second derivative of the phase as $\overline\varphi''_{\xi} =\varphi''_{\xi}/\alpha\beta\rho''$ with its form evident from (\ref{eq:quad-2dir0-y}) and where the phases are given by (\ref{eq:quad-2ph0-y}) and their second derivatives were computed in the form (\ref{eq:quad-2dir0-y}).

The square of the Fourier-transformed amplitude of the EM field, ${\cal A}_{\tt out}^2({\vec x}_i,{\vec x}'')$ ij the region outside the astroid caustic that is needed in (\ref{eq:pow-blur}), may be given as
{}
\begin{eqnarray}
{\cal A}_{\tt out}^2({\vec x}_i,{\vec x}'')&=&
 \frac{1}{2\pi\alpha\beta\rho''}{\cal B}_{\tt out}^2({\vec x}_i,{\vec x}''),
  \label{eq:psf-quad_out-AB}
\end{eqnarray}
with ${\cal B}_{\tt out}^2({\vec x}_i,{\vec x}'')$ up the terms of the oder of ${\cal O}\big(({4\beta_2}/{\alpha\beta\rho''})^3\big)$ given as below
{}
\begin{eqnarray}
{\cal B}_{\tt out}^2({\vec x}_i,{\vec x}'')&=&
\Big( {\cal A}^2_0+ {\cal A}^2_\pi\Big)
\Big( 1+{\textstyle\frac{1}{16}}\Big(1+15\cos[4(\phi''-\phi_s)]\Big)\Big(\frac{4\beta_2}{\alpha\beta\rho''}\Big)^2  \Big)+
\Big( {\cal A}^2_\pi- {\cal A}^2_0\Big)\cos[2(\phi''-\phi_s)]\Big(\frac{4\beta_2}{\alpha\beta\rho''}\Big)+\nonumber\\[4pt]
 &&\hskip -30pt +\,
 2{\cal A}_0{\cal A}_{\pi}
\sin\Big[2\alpha\beta\rho''\Big(1+  {\textstyle\frac{3}{8}}
\sin^2[2(\phi''-\phi_s)]\Big(\frac{4\beta_2}{\alpha\beta\rho''}\Big)^2\Big)\Big]
\Big(
 1- {\textstyle\frac{1}{16}}\Big(3-11\cos[4(\phi''-\phi_s)]\Big) \Big(\frac{4\beta_2}{\alpha\beta\rho''}\Big)^2 \Big).
  \label{eq:psf-quad_out-AB2}
\end{eqnarray}

With this result, we may now proceed with evaluating the optical properties of the quadrupole SGL.

\section{Intensity distribution and signal received by a telescope}
\label{sec:intense-in}

Expression (\ref{eq:pow-blur}) allows us to compute the power received from the resolved source. For an actual astrophysical source, $B_s({\vec x}')$ is, of course, an arbitrary function of the coordinates ${\vec x}'$ and thus the integral can only be evaluated numerically. However, we can obtain an analytic result in the simple case of a disk of uniform brightness, characterized by ${B}_s({\vec x}') ={B}_s$. In this case, the integral (\ref{eq:pow-blur}) takes the form
  {}
\begin{eqnarray}
I_{\tt }({\vec x}_i,{\vec x}_0) &=&
 \mu_0\frac{B_{\tt s}}{z^2_0}  \Big(\frac{kd^2}{8f}\Big)^2\hskip-4pt
\iint d^2{\vec x}''  {\cal A}^2({\vec x}_i,{\vec x}''),
  \label{eq:intense-y}
\end{eqnarray}
where the integration over $d^2{\vec x}'' $ is done within the image boundary, with respect to the telescope position ${\vec x}_0$.

Expression (\ref{eq:intense-y}) is a good choice when dealing with the integral form for the EM field amplitude as given by  (\ref{eq:amp-quad}). Using this expression, it is possible to separate regions inside and outside the caustic boundary, and obtain a finite result. However, with ${\cal A}^2({\vec x}_i,{\vec x}'')$ given by (\ref{eq:amp-quad}), this integral is highly oscillatory, difficult to evaluate computationally.

To tame the behavior of the integral, we may use the quartic solution for ${\cal A}^2({\vec x}_i,{\vec x}'')$ presented in \cite{Turyshev-Toth:2021-quartic}. The quartic solution is a good approximation to the integral  (\ref{eq:amp-quad})
everywhere except for the immediate vicinity of the caustic boundary. Therefore, the integration in (\ref{eq:intense-y}) needs to account for that as was done in \cite{Turyshev-Toth:2021-quartic}. To do that, we will rely on the two approximate solutions for ${\cal A}^2({\vec x}_i,{\vec x}'')$ that we developed for the inner part of the caustic, ${\cal A}_{\tt in}^2({\vec x}_i,{\vec x}'')$, and the outer part, ${\cal A}_{\tt out}^2({\vec x}_i,{\vec x}'')$,  which are respectively given by (\ref{eq:psf-quad_in-AB}) and (\ref{eq:psf-quad_out-AB}).

To properly identify integration limits in (\ref{eq:intense-y}), we need to take into account all possible positions of the astroid with respect to the caustic boundary. These include the case when
\begin{inparaenum}[i)]
\item astroid caustic is completely within the image;
\item the center of the caustic is still within the image, but some parts of it are outside;
\item the center of the caustic is outside the image, but some parts of it are still within the image, and
\item the entire caustic is outside the image.
\end{inparaenum}
The first pair of these four cases correspond to the telescope being positioned within the image nominally projected by the monopole PSF, while the second pair represent cases when the telescope is outside. For details, see \cite{Turyshev-Toth:2020-photom,Turyshev-Toth:2020-extend,Turyshev-Toth:2022-broad-SNR,Turyshev-Toth:2022-mono-SNR}.

The first of these four cases is shown in Fig.~\ref{fig:pos-caustic}~(left). The second and third cases are characterized by additional parameters that are shown in Fig.~\ref{fig:pos-caustic}~(right). We now consider these cases in detail, starting with the telescope fully inside the projected image of the source.

\subsection{Intensity distribution for the telescope within the image}
\label{sec:blur-in}

To integrate (\ref{eq:intense-y}), we introduce a new coordinate system in the source plane, ${\vec x}''$, with the origin at the telescope position:
${\vec x}'-{\vec x}'_0={\vec x}''$. As the vector ${\vec x}'_0$ is constant, $dx'dy'=dx''dy''$. Switching to polar coordinates, $(x'',y'')\rightarrow (r'',\phi'')$, we can represent the circular edge of the source, $R_\oplus$, by the curve $\rho_\oplus(\phi'')$ that is given by the 
relation (see  \cite{Turyshev-Toth:2020-extend}):
{}
\begin{eqnarray}
\rho_\oplus(\phi'')
&=&\sqrt{R_\oplus^2-{\rho'_0}^2\sin^2\phi''}-\rho'_0\cos\phi''\equiv
R_\oplus\Big(\sqrt{1-\Big(\frac{\rho_0}{r_\oplus}\Big)^2\sin^2\phi''}-\frac{\rho_0}{r_\oplus}\cos\phi''\Big),
\label{eq:rho+y}
\end{eqnarray}
where $\rho_0$ is the telescope's position within the image (note that $\rho_0'/R_\oplus \equiv \rho_0/r_\oplus$).

From (\ref{eq:caust}) this astroid caustic's boundary, projected onto the source plane, has the form
{}
\begin{eqnarray}
\rho_{\tt ac}(\phi'')&=&\Big(\frac{4\beta_2}{\alpha\beta}\Big)\Big(\sin^\frac{2}{3}
(\phi''+\phi_0-\phi_s)
+\cos^\frac{2}{3}(\phi''+\phi_0-\phi_s)
\Big)^{-\frac{3}{2}}.
  \label{eq:caust-eps-y2}
\end{eqnarray}

Expressions (\ref{eq:rho+y}) and (\ref{eq:caust-eps-y2}) can be used to establish the relationships between the boundary of the astroid caustic, $\rho_{\tt ac}(\phi'')$, and that of the image of the source, $\rho_\oplus(\phi'')$.

\subsubsection{Astroid caustic is entirely within the image}

Consider the first case, when the astroid caustic is completely inside the image boundary. To develop insight into the behavior of the signal received by the telescope, we use the results for ${\cal A}_{\tt in}^2({\vec x}_i,{\vec x}'')$ and ${\cal A}_{\tt out}^2({\vec x}_i,{\vec x}'')$ developed earlier and given by (\ref{eq:psf-quad_in-AB}) (with $ {\cal B}_{\tt in}^2({\vec x}_i,{\vec x}'')$ from (\ref{eq:psf-quad_in-AB2})) and (\ref{eq:psf-quad_out-AB})--(\ref{eq:psf-quad_out-AB2}), correspondingly. From (\ref{eq:intense-y}), these results yield the following expression for the intensity in the telescope focal plane:
  {}
\begin{eqnarray}
I_{\tt }({\vec x}_i,{\vec x}_0) &=&
 \mu_0\frac{B_{\tt s}}{z^2_0} \Big(\frac{kd^2}{8f}\Big)^2  \bigg\{ \int_0^{2\pi} \hskip -8pt d\phi'' \int_{0}^{\rho_{\tt ac}}\hskip -4pt \rho''d\rho''  {\cal A}_{\tt in}^2({\vec x}_i,{\vec x}'')+ \int_0^{2\pi} \hskip -8pt d\phi'' \int_{\rho_{\tt ac}}^{\rho_\oplus}\hskip -4pt \rho''d\rho''  {\cal A}_{\tt out}^2({\vec x}_i,{\vec x}'')\bigg\}.~~~
  \label{eq:intense-y-qu-ap-y*}
\end{eqnarray}
After rearranging the integration limits and using expressions for ${\cal A}_{\tt in}^2({\vec x}_i,{\vec x}'')$ and ${\cal A}_{\tt out}^2({\vec x}_i,{\vec x}'')$  and given by (\ref{eq:psf-quad_in-AB}) and (\ref{eq:psf-quad_out-AB}), correspondingly, this expression may be presented in the following, equivalent form:
  {}
\begin{eqnarray}
I_{\tt }({\vec x}_i,{\vec x}_0) &=&
 \mu_0\frac{B_{\tt s}}{z^2_0} \Big(\frac{kd^2}{8f}\Big)^2   \frac{1}{\alpha\beta}\bigg\{  \frac{1}{2\pi}\int_0^{2\pi} \hskip -8pt d\phi'' \int_0^{\rho_\oplus}\hskip -4pt d\rho''  {\cal B}_{\tt out}^2({\vec x}_i,{\vec x}'')- \nonumber\\
 &&\hskip 80pt -\,
\frac{1}{2\pi} \int_0^{2\pi} \hskip -8pt d\phi'' \int_{0}^{\rho_{\tt ac}}\hskip -4pt d\rho''  \Big(
 {\cal B}_{\tt out}^2({\vec x}_i,{\vec x}'')- \Big(\frac{\alpha\beta\rho''}{4\beta_2}\Big){\cal B}_{\tt in}^2({\vec x}_i,{\vec x}'')\Big)\bigg\}.~~~
  \label{eq:intense-y-qu-ap-y}
\end{eqnarray}

We recognize the first term in the expression as the intensity of light received in the focal plane of an imaging telescope in the case of the monopole PSF, i.e., treating the sun as a spherically-symmetric lens \cite{Turyshev-Toth:2020-photom,Turyshev-Toth:2020-extend,Turyshev-Toth:2022-broad-SNR,Turyshev-Toth:2022-mono-SNR}. To evaluate the remaining terms in (\ref{eq:intense-y-qu-ap-y}), we observe that ${\cal B}_{\tt out}^2({\vec x}_i,{\vec x}'')\simeq 1+{\cal O}(y)$ and ${\cal B}_{\tt in}^2({\vec x}_i,{\vec x}'')\simeq 1+{\cal O}(x)$. With the numerical values for $\alpha$ and  $\beta_2$ from (\ref{eq:zerJ}) and (\ref{eq:beta2}), correspondingly, we see that in the inner part of the caustic, the ratio $({\alpha\beta\rho''})/({4\beta_2})\equiv x<1$; see (\ref{eq:quad-small-par}). Therefore, the second integral in (\ref{eq:intense-y-qu-ap-y}) represents a reduction of the signal estimated based on the monopole PSF by accounting for the fact that the Sun is an extended oblate object.

Looking at the functional form of ${\cal B}_{\tt in}^2({\vec x}_i,{\vec x}'')$ and ${\cal B}_{\tt out}^2({\vec x}_i,{\vec x}'')$ we see that the second term in (\ref{eq:intense-y-qu-ap-y}) is not uniform, but has azimuthal dependence because of the functional form of the caustic boundary itself, $\rho_{\tt ac}(\phi'')$, given by (\ref{eq:caust-eps-y2}) and also the azimuthal dependence present in ${\cal B}_{\tt in}^2({\vec x}_i,{\vec x}'')$. So, most of the signal reaches the telescope's focal plane, but due to the presence of second term in (\ref{eq:intense-y-qu-ap-y}), this  signal is scrambled.

\begin{figure}[t!]
\includegraphics[scale=0.75]{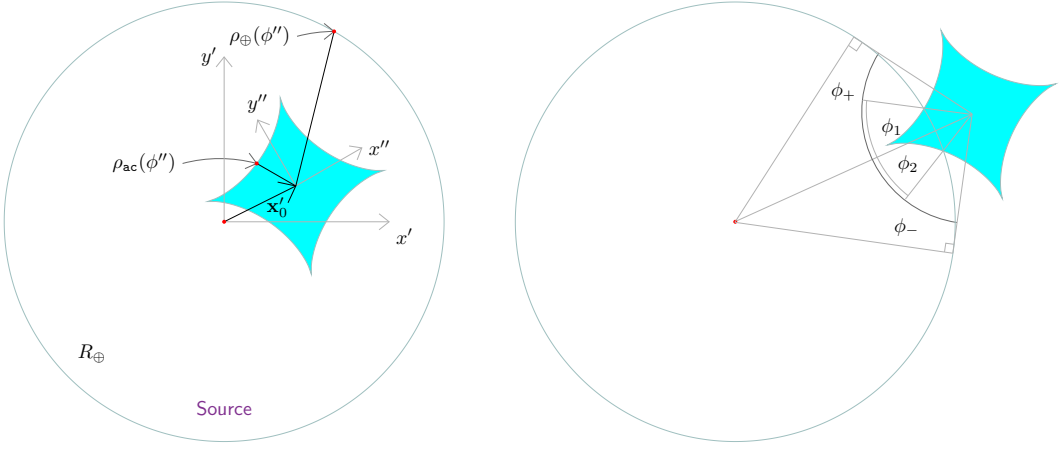}
\caption{\label{fig:pos-caustic} Geometry and parametrization of the extended SGL for imaging purposes. Left: the astroid caustic projection on the source plane is inside the source. Right: When the caustic projection boundary intersects the source boundary, it is characterized by the angles $\phi_{1,2}$. Furthermore, if the center of the caustic projection is outside the source, it is characterized by $\phi_\pm$. Otherwise, the parametrization is the same as for the caustic projection that's completely inside the source. Additional configurations, including the caustic projection being entirely outside the source, or a caustic with two or three cusps inside the source and the center either inside or outside can be treated similarly, with additional angular parameters characterizing intersections.
}
\end{figure}

To study the signal from a source in the case on the extended PSF, we evaluate the power received at the telescope focal plane. For that, we need to integrate the intensity in the focal plane over the area occupied by the Einstein ring. Similarly to the discussion in \cite{Turyshev-Toth:2020-extend}, one can estimate this quantity as
{}
\begin{eqnarray}
P_{\tt fp}({\vec x}_0)&=&\epsilon_{\tt ee}P_{\tt }({\vec x}_0)\equiv \epsilon_{\tt ee} \int^{2\pi}_0 \hskip -4pt  d\phi_i \int_0^\infty
\hskip 0pt I_{\tt }({\vec x}_i,{\vec x}_0) \rho_i d\rho_i,
  \label{eq:pow-frac-bl}
\end{eqnarray}
where $I_{\tt }({\vec x}_i,{\vec x}_0) $ is from (\ref{eq:intense-y-qu-ap-y}) and also we introduce the encircled energy factor, $\epsilon_{\tt ee}=0.69$.

To evaluate the integral in (\ref{eq:pow-frac-bl}), we notice that only two quantities in that expression depend on $\rho_i$ and $\phi_i$, namely ${\cal B}_{\tt out}^2({\vec x}_i,{\vec x}'')$ and ${\cal B}_{\tt in}^2({\vec x}_i,{\vec x}'')$. As was discussed in \cite{Turyshev-Toth:2020-extend,Turyshev-Toth:2022-morphology-ER}, for any set of stationary phase solutions, $\phi^{[j]}$, where $j=\in [1,4]$ for the quadrupole PSF and $j=\in [1,2]$ for the monopole PSF, the following two useful relations exist:
{}
\begin{eqnarray}
\int^{2\pi}_0 \hskip -4pt  d\phi_i \int_0^\infty
\hskip 0pt  \rho_i d\rho_i   \Big(\frac{2J_1(\alpha d\, \hat u(\phi^{[j]}_\xi, \vec x_i))}{\alpha d\, \hat u(\phi^{[j]},\vec x_i)}\Big)^2&=&
4\pi \Big(\frac{\lambda f}{\pi d}\Big)^2,
  \label{eq:int-fp1*}\\
\int^{2\pi}_0 \hskip -4pt  d\phi_i \int_0^\infty
\hskip 0pt  \rho_i d\rho_i  \Big(\frac{2J_1(\alpha d\, \hat u(\phi^{[j]}_\xi, \vec x_i))}{\alpha d\, \hat u(\phi^{[j]},\vec x_i)}\Big)\Big(\frac{2J_1(\alpha d\, \hat u(\phi^{[k]}_\xi, \vec x_i))}{\alpha d\, \hat u(\phi^{[k]},\vec x_i)}\Big)&\lesssim&
5\times 10^{-3} \cdot 4\pi \Big(\frac{\lambda f}{\pi d}\Big)^2.
  \label{eq:int-fp2*}
\end{eqnarray}

Applying these relations on ${\cal B}_{\tt out}^2({\vec x}_i,{\vec x}'')$ and ${\cal B}_{\tt in}^2({\vec x}_i,{\vec x}'')$ given by   (\ref{eq:psf-quad_in-AB2}) and (\ref{eq:psf-quad_out-AB2}), correspondingly, we have
{}
\begin{eqnarray}
\int^{2\pi}_0 \hskip -4pt  d\phi_i \int_0^\infty
\hskip 0pt  \rho_i d\rho_i  {\cal B}_{\tt out}^2({\vec x}_i,{\vec x}'')&=&
8\pi \Big(\frac{\lambda f}{\pi d}\Big)^2 \Big(1+ {\cal O}(y^2) \Big),
  \label{eq:int-B1}\\
\int^{2\pi}_0 \hskip -4pt  d\phi_i \int_0^\infty
\hskip 0pt  \rho_i d\rho_i  {\cal B}_{\tt in}^2({\vec x}_i,{\vec x}'')&=& \nonumber\\
&&\hskip -100pt\,=
16\pi \Big(\frac{\lambda f}{\pi d}\Big)^2 \bigg\{1+ \Big(\frac{\alpha\beta\rho''}{4\beta_2}\Big)^2+
 {\textstyle\frac{1}{4}}\Big(9-5\cos[4(\phi''+\phi_0-\phi_s)]\Big)\Big(\frac{\alpha\beta\rho''}{4\beta_2}\Big)^4+{\cal O}\Big(\Big(\frac{\alpha\beta\rho''}{4\beta_2}\Big)^6\Big) \bigg\},
  \label{eq:int-B2}
\end{eqnarray}
where we neglected the mixed terms of the type ${\cal A}_j{\cal A}_k$ and also kept (\ref{eq:int-B1}) to the order of ${\cal O}(y^2) $ as even the terms of that order are already small and rapidly diminishing.

With results (\ref{eq:int-B1})--(\ref{eq:int-B2}), expression (\ref{eq:pow-frac-bl})  for the power received at the telescope's focal plane takes the form:
  {}
\begin{eqnarray}
P_{\tt fp}({\vec x}_0)&=& \epsilon_{\tt ee} \int^{2\pi}_0 \hskip -4pt  d\phi_i \int_0^\infty
\hskip 0pt I_{\tt }({\vec x}_i,{\vec x}_0) \rho_i d\rho_i=\nonumber\\
&&\hskip -40pt =\,
\epsilon_{\tt ee} \mu_0\frac{B_{\tt s}}{z^2_0}\frac{\pi d^2 R_\oplus}{2\alpha\beta}\bigg\{  \epsilon(\rho_0) -
 \nonumber\\
 &&\hskip -20pt -\,
\frac{1}{2\pi R_\oplus} \int_0^{2\pi} \hskip -8pt d\phi''  \Big(
\rho_{\tt ac}- \Big(\frac{\alpha\beta}{4\beta_2}\Big)\Big(\rho_{\tt ac}^2+ {\textstyle\frac{1}{2}}\Big(\frac{\alpha\beta}{4\beta_2}\Big)^2\rho_{\tt ac}^4+
 {\textstyle\frac{1}{12}}\Big(9-5\cos[4(\phi''+\phi_0-\phi_s)]\Big)\Big(\frac{\alpha\beta}{4\beta_2}\Big)^4\rho_{\tt ac}^6\Big)\bigg\},
 \label{eq:power0}
\end{eqnarray}
where the blur factor $\epsilon(\rho_0)$  is given by the expression \cite{Turyshev-Toth:2020-photom}:
{}
\begin{eqnarray}
\epsilon(\rho_0)
&=&\frac{1}{2\pi}\int_0^{2\pi} \hskip -3pt d\phi''\sqrt{1-\Big(\frac{\rho_0}{r_\oplus}\Big)^2\sin^2\phi''}=\frac{2}{\pi}{\tt E}\Big[\Big(\frac{\rho_0}{r_\oplus}\Big)^2\Big],
\label{eq:eps_r0}
\end{eqnarray}
where ${\tt E}[x]$ is the elliptic integral \cite{Abramovitz-Stegun:1965,Korn-Korn:1968}.
Using the expression for the astroid caustic $\rho_{\tt ac}(\phi'')$ given by (\ref{eq:caust-eps-y2}) we define quantity $q(\phi'')$ as below
{}
\begin{eqnarray}
\rho_{\tt ac}(\phi'')&=&\Big(\frac{4\beta_2}{\alpha\beta}\Big)q(\phi''), \qquad {\rm where} \qquad
q(\phi'')=\Big(\sin^\frac{2}{3}
(\phi''+\phi_0-\phi_s)
+\cos^\frac{2}{3}(\phi''+\phi_0-\phi_s)\Big)^{-\frac{3}{2}}.
  \label{eq:caust-eps-y2-q}
\end{eqnarray}
With this definition, we take the integrals over $d\phi''$ in (\ref{eq:power0}), that results in
{}
\begin{eqnarray}
P_{\tt fp}({\vec x}_0)&=&
\epsilon_{\tt ee} \mu_0\frac{B_{\tt s}}{z^2_0}\frac{\pi d^2 R_\oplus}{2\alpha\beta}\bigg\{  \epsilon(\rho_0) -
\nonumber\\
 &-&
\Big(\frac{4\beta_2}{\alpha r_\oplus}\Big) \frac{1}{2\pi} \int_0^{2\pi} \hskip -8pt d\phi''  \bigg\{
q(\phi'')- q^{2}(\phi'')- {\textstyle\frac{1}{2}}q^{4}(\phi'')-
{\textstyle\frac{1}{12}}\Big(9-5\cos[4(\phi''+\phi_0-\phi_s)]\Big)
 q^{6}(\phi'')\bigg\}\bigg\}
=\nonumber\\
&=&
\epsilon_{\tt ee} \mu_0\frac{B_{\tt s}}{z^2_0}\frac{\pi d^2 R_\oplus}{2\alpha\beta}\bigg\{  \epsilon(\rho_0) -
 0.106
\Big(\frac{4\beta_2}{\alpha r_\oplus}\Big)\bigg\}.~~~
  \label{eq:power00}
\end{eqnarray}

From (\ref{eq:4beta2}), we estimate
\begin{eqnarray}
\frac{4\beta_2}{\alpha r_\oplus}&=&\frac{2J_2 \Big(\frac{R^2_\odot }{\sqrt{2r_g\bar z}}\Big)\sin^2\beta_s}{({\overline z}/z_0)R_\oplus}=0.429\, \sin^2\beta_s\Big(\frac{650~{\rm AU}}{\overline z}\Big)^\frac{3}{2}\Big(\frac{z_0}{30~{\rm pc}}\Big)\Big(\frac{6378\,{\rm km}}{R_\oplus}\Big).
  \label{eq:4beta22}
  \end{eqnarray}
As a result, the fraction of light by which the presence of the solar quadrupole reduces the amount of light received within the image compared to the monopole case is given as
{}
\begin{eqnarray}
&&
 0.106\Big( \frac{4\beta_2}{\alpha r_\oplus}\Big)\simeq
0.045\, \sin^2\beta_s\Big(\frac{650~{\rm AU}}{\overline z}\Big)^\frac{3}{2}\Big(\frac{z_0}{30~{\rm pc}}\Big)\Big(\frac{6378\,{\rm km}}{R_\oplus}\Big).
  \label{eq:fraction}
  \end{eqnarray}

Expression (\ref{eq:fraction}) suggest that in the case of the quadrupole PSF some fraction of the signal photons may fall outside the source image, as it defined by the monopole PSF. All these photons are still in the image proximity but are scrambled. So, the signal is there, but to recover the image, the signal must be unscrambled, i.e., deconvolved.

\subsubsection{Center of the astroid is still within the image, but some parts of it are out}

In the vicinity of the image boundary, some part of the astroid caustic extends beyond the image boundary, so for some angles $\phi''$ the parametric equation for the caustic (\ref{eq:caust}) intersect with that of the image boundary (\ref{eq:rho+y}), setting the condition to determine the angles (up to eight) at which such an intersection occurs:
  {}
\begin{eqnarray}
\rho_{\tt ac}(\phi'')=\rho_\oplus(\phi'') \qquad \rightarrow \qquad \phi''=\Big\{\phi_1,\phi_2,\phi_3,\phi_4, \phi_5,\phi_6,\phi_7,\phi_8\Big\}.
  \label{eq:bound_in}
\end{eqnarray}
Note that there are eight angles only when all four cusps are outside the image boundary while the center of the astroid remains inside. (This presupposes that the astroid is comparable in size and area to the projected image.) Here, for simplicity, we assume that the center of the astroid caustic is inside the image and only one cusp is outside the image area.

Assuming that that are only two of such angles $(\phi_1,\phi_2)\equiv(\phi_1(\vec x_0),\phi_2(\vec x_0))$ (see also Fig.~\ref{fig:pos-caustic}), the integration for the intensity in the telescope's focal plane in (\ref{eq:intense-y}) changes to
  {}
\begin{eqnarray}
I_{\tt }({\vec x}_i,{\vec x}_0) &=&
 \mu_0\frac{B_{\tt s}}{z^2_0} \Big(\frac{kd^2}{8f}\Big)^2  \bigg\{ \int_{\phi_2}^{2\pi-\phi_1} \hskip -8pt d\phi'' \int_{0}^{\rho_{\tt ac}}\hskip -4pt \rho''d\rho''  {\cal A}_{\tt in}^2({\vec x}_i,{\vec x}'')
 +
 \int_{-\phi_1}^{\phi_2} \hskip -8pt d\phi'' \int_{0}^{\rho_\oplus}\hskip -4pt \rho''d\rho''  {\cal A}_{\tt in}^2({\vec x}_i,{\vec x}'')
 + \nonumber\\
 &&\hskip 80pt \,+
 \int_{\phi_2}^{2\pi-\phi_1} \hskip -8pt d\phi'' \int_{\rho_{\tt ac}}^{\rho_\oplus}\hskip -4pt \rho''d\rho''  {\cal A}_{\tt out}^2({\vec x}_i,{\vec x}'')\bigg\}.~~~
  \label{eq:intense-out*}
\end{eqnarray}
Similarly to (\ref{eq:intense-y-qu-ap-y}), after rearranging the integration limits and using expressions for ${\cal A}_{\tt in}^2({\vec x}_i,{\vec x}'')$ and ${\cal A}_{\tt out}^2({\vec x}_i,{\vec x}'')$  and given by (\ref{eq:psf-quad_in-AB}) and (\ref{eq:psf-quad_out-AB}), correspondingly, this expression may be presented in the following equivalent form:
  {}
\begin{eqnarray}
I_{\tt }({\vec x}_i,{\vec x}_0) &=&
 \mu_0\frac{B_{\tt s}}{z^2_0} \Big(\frac{kd^2}{8f}\Big)^2   \frac{1}{\alpha\beta}\times\nonumber\\
&&\hskip -40pt \times\,
\bigg\{  \frac{1}{2\pi}\int_0^{2\pi} \hskip -8pt d\phi'' \int_0^{\rho_\oplus}\hskip -4pt d\rho''  {\cal B}_{\tt out}^2({\vec x}_i,{\vec x}'')-
 \frac{1}{2\pi} \int_0^{2\pi} \hskip -8pt d\phi'' \int_{0}^{\rho_{\tt ac}}\hskip -4pt d\rho''  \Big(
 {\cal B}_{\tt out}^2({\vec x}_i,{\vec x}'')- \Big(\frac{\alpha\beta\rho''}{4\beta_2}\Big){\cal B}_{\tt in}^2({\vec x}_i,{\vec x}'')\Big)+
 \nonumber\\
 &&\hskip 70pt +\,
\frac{1}{2\pi} \int_{-\phi_1}^{\phi_2} \hskip -8pt d\phi'' \int_{\rho_\oplus}^{\rho_{\tt ac}}\hskip -4pt d\rho''  \Big(
 {\cal B}_{\tt out}^2({\vec x}_i,{\vec x}'')- \Big(\frac{\alpha\beta\rho''}{4\beta_2}\Big){\cal B}_{\tt in}^2({\vec x}_i,{\vec x}'')\Big)\bigg\}.~~~
  \label{eq:intense-out*2}
\end{eqnarray}

By looking at (\ref{eq:intense-out*2}), we can immediately see that the first two integrals in this expression are identical to those in (\ref{eq:intense-y-qu-ap-y}) and represent a reduction of the signals due to the PSF of the oblate Sun. The last term is new, implying that some part of the reduced signal is recovered. The best way to evaluate this behavior is to study the power received in the telescope's focal plane.  For that, following the logic used to develop (\ref{eq:power0}), we rely on (\ref{eq:int-fp1*}) and (\ref{eq:int-fp2*}) to derive
  {}
\begin{eqnarray}
P_{\tt fp}({\vec x}_0)&=& \epsilon_{\tt ee} \int^{2\pi}_0 \hskip -4pt  d\phi_i \int_0^\infty
\hskip 0pt I_{\tt }({\vec x}_i,{\vec x}_0) \rho_i d\rho_i=
\epsilon_{\tt ee} \mu_0\frac{B_{\tt s}}{z^2_0}\frac{\pi d^2 R_\oplus}{2\alpha\beta}\bigg\{  \epsilon(\rho_0)  -
 0.106
\Big(\frac{4\beta_2}{\alpha r_\oplus}\Big)+\nonumber\\
&&\hskip 10pt  +\,
\frac{1}{2\pi R_\oplus} \int_{-\phi_1}^{\phi_2} \hskip -8pt d\phi''  \Big(
\rho_{\tt ac}-\rho_\oplus- \Big(\frac{\alpha\beta}{4\beta_2}\Big)\Big(\rho_{\tt ac}^2-\rho_\oplus^2+ {\textstyle\frac{1}{2}}\Big(\frac{\alpha\beta}{4\beta_2}\Big)^2(\rho_{\tt ac}^4-\rho_\oplus^4)+\nonumber\\
&&\hskip 100pt  +\,
 {\textstyle\frac{1}{12}}\Big(9-5\cos[4(\phi''+\phi_0-\phi_s)]\Big)\Big(\frac{\alpha\beta}{4\beta_2}\Big)^4(\rho_{\tt ac}^6-\rho_\oplus^6)
\Big)\Big)
\bigg\}.
  \label{eq:power02+}
\end{eqnarray}

We can simplify this integral in the limiting case when $\phi_1,\phi_2,\phi_0 \ll 1^\circ$ are all small. That allows as to set 
the values of $\rho_0$ and $\rho_\oplus$
at $\phi''=0$. For that, from (\ref{eq:rho+y}) and (\ref{eq:caust-eps-y2}) we have  the following relevant values (with $q(\phi'')$ from (\ref{eq:caust-eps-y2-q})):
{}
\begin{eqnarray}
\rho_\oplus(0)
&=&R_\oplus\Big(1-\frac{\rho_0}{r_\oplus}\Big) \qquad {\rm and} \qquad
\rho_{\tt ac}(0)=
\Big(\frac{4\beta_2}{\alpha\beta}\Big) q(\phi_0),
  \label{eq:at-zero}
\end{eqnarray}
which yields
{}
\begin{eqnarray}
&&
\frac{1}{2\pi R_\oplus} \int_{-\phi_1}^{\phi_2} \hskip -8pt d\phi''  \Big(
\rho_{\tt ac}-\rho_\oplus- \Big(\frac{\alpha\beta}{4\beta_2}\Big)\Big(\rho_{\tt ac}^2-\rho_\oplus^2+ {\textstyle\frac{1}{2}}\Big(\frac{\alpha\beta}{4\beta_2}\Big)^2(\rho_{\tt ac}^4-\rho_\oplus^4)+\nonumber\\
&&\hskip 107pt  +\,
 {\textstyle\frac{1}{12}}\Big(9-5\cos[4(\phi''+\phi_0-\phi_s)]\Big)\Big(\frac{\alpha\beta}{4\beta_2}\Big)^4(\rho_{\tt ac}^6-\rho_\oplus^6)
\Big)\Big)
\bigg\}=\nonumber\\
&=&\frac{\phi_1+\phi_2}{2\pi}\bigg\{\Big(\frac{\alpha r_\oplus}{4\beta_2}\Big)\Big(1-\frac{\rho_0}{r_\oplus}\Big)^2\bigg(1+{\textstyle\frac{1}{2}}\Big(\frac{\alpha r_\oplus}{4\beta_2}\Big)^2\Big(1-\frac{\rho_0}{r_\oplus}\Big)^2+
 {\textstyle\frac{1}{12}}\Big(9-5\cos[4(\phi_0-\phi_s)]\Big)\Big(\frac{\alpha r_\oplus}{4\beta_2}\Big)^4\Big(1-\frac{\rho_0}{r_\oplus}\Big)^4
\bigg)-
\nonumber\\&&\hskip 40pt-\,
\Big(1-\frac{\rho_0}{r_\oplus}\Big)+
\Big(\frac{4\beta_2}{\alpha r_\oplus}\Big)\Big(q(\phi_0)-q^2(\phi_0)
-{\textstyle\frac{1}{2}}q^4(\phi_0)- {\textstyle\frac{1}{12}}\Big(9-5\cos[4(\phi_0-\phi_s)]\Big)q^6(\phi_0)\Big)
\bigg\}\leq\nonumber\\
&&\hskip 20pt  \leq\,
-\frac{\phi_1+\phi_2}{2\pi}\bigg\{
\frac{\delta\rho_0}{r_\oplus}+{\textstyle\frac{5}{6}}
\Big(\frac{4\beta_2}{\alpha r_\oplus}\Big)+{\cal O}\Big(\Big(\frac{\delta\rho_0}{r_\oplus}\Big)^2\Big)
\bigg\},~~~~~
  \label{eq:power-int2}
\end{eqnarray}
where $\delta \rho_0=r_\oplus-\rho_0$.
Remember that in this case $\rho_0<r_\oplus$, thus the expression (\ref{eq:power-int2}) is positive.

Evaluating this result similarly to (\ref{eq:power00}), we see that in this area near the image boundary, depending on the telescope position, $\rho_0$, some photons are being further removed from the area enveloping the image. Thus, we can see that the flux anticipated from a target is  $\sim 6\%$ weaker than in the monopole case, as shown by (\ref{eq:fraction}).

\subsection{Intensity distribution at an off-image telescope position}
\label{sec:extend-off-image}

The same conditions to derive (\ref{eq:intense-y-qu-ap-y}) are valid where the telescope is at an off-image position, so the power received by the telescope takes the same form. The only difference comes from the fact that we are outside the image, thus, the integration limits change. First, we note that the circular edge of the source, $R_\oplus$, is given by a curve, $\rho_\oplus(\phi'')$, the radial distance of which in this polar coordinate system is given as
{}
\begin{eqnarray}
\rho^\pm_\oplus(\phi'') &=&\pm \sqrt{R_\oplus^2-{\rho'_0}^2\sin^2\phi''}+\rho'_0\cos\phi''=R_\oplus\Big(\pm\sqrt{1-\Big(\frac{\rho_0}{r_\oplus}\Big)^2\sin^2\phi''}+\frac{\rho_0}{r_\oplus}\cos\phi''\Big),
\label{eq:rho++B}
\end{eqnarray}
with the angle $\phi''$ in this case is defined so that $\phi''=0$ when pointing at the center of the source. The angle $\phi''$ varies only within the range $\phi''\in [\phi_-,\phi_+]$, with $\phi_\pm=\pm \arcsin ({R_\oplus}/{\rho'_0})$. Given the sign in front of the square root in (\ref{eq:rho++B}), for any angle $\phi''$ there will be two solutions for $\rho_\oplus(\phi'')$, given as $(\rho^-_\oplus,\rho^+_\oplus)$.

\subsubsection{Off-image, caustic does not intersect the image boundary}

Assuming that the brightness of the source in this region is uniform, $B(x',y')=B_{\tt s}$, we use (\ref{eq:rho++B}) and evaluate (\ref{eq:pow-blur}) for this set of conditions:
 {}
\begin{eqnarray}
I_{\tt off}({\vec x}_i,{\vec x}_0)  &=&  \mu_0\frac{B_{\tt s}}{z^2_0} \Big(\frac{kd^2}{8f}\Big)^2 \int_{\phi_-}^{\phi_+} \hskip -8pt d\phi''
\int_{\rho^-_\oplus}^{\rho^+_\oplus}\hskip -3pt \rho'' d\rho''
  {\cal A}^2_{\tt out}({\vec x}_i,{\vec x}'')=
\mu_0\frac{B_{\tt s}}{z^2_0}  \Big(\frac{kd^2}{8f}\Big)^2    \frac{ 1}{2\pi \alpha\beta} \int_{\phi_-}^{\phi_+} \hskip -8pt d\phi''
\int_{\rho^-_\oplus}^{\rho^+_\oplus}\hskip -3pt d\rho''
  {\cal B}^2_{\tt out}({\vec x}_i,{\vec x}'').~~~~
  \label{eq:intense-off-quad}
\end{eqnarray}
We next compute the power deposited at off image locations as
  {}
\begin{eqnarray}
P_{\tt fp.out}({\vec x}_0)&=& \epsilon_{\tt ee} \int^{2\pi}_0 \hskip -4pt  d\phi_i \int_0^\infty
\hskip 0pt I_{\tt off}({\vec x}_i,{\vec x}_0) \rho_i d\rho_i=
\epsilon_{\tt ee} \mu_0\frac{B_{\tt s}}{z^2_0}\frac{\pi d^2}{2\alpha\beta} \frac{1}{2\pi} \int_{\phi_-}^{\phi_+} \hskip -8pt d\phi''
\int_{\rho^-_\oplus}^{\rho^+_\oplus}\hskip -3pt d\rho''
= \epsilon_{\tt ee} \mu_0\frac{B_{\tt s}}{z^2_0}\frac{\pi d^2 R_\oplus}{2\alpha\beta} \beta(\rho_0),
  \label{eq:power0+off}
\end{eqnarray}
where  the factor $\beta (\rho_0)$ given by the following expression:
{}
\begin{eqnarray}
\beta (\rho_0)
&=&\frac{1}{\pi}\int_{\phi_-}^{\phi_+} \hskip -3pt d\phi''\sqrt{1-\Big(\frac{\rho_0}{r_\oplus}\Big)^2\sin^2\phi''}=\frac{2}{\pi}{\tt E}\Big[\arcsin \frac{r_\oplus}{\rho_0},\Big(\frac{\rho_0}{r_\oplus}\Big)^2\Big],
\label{eq:beta_r0}
\end{eqnarray}
where ${\tt E}[a,x]$ is the incomplete elliptic integral \cite{Abramovitz-Stegun:1965}. This result is identical to that obtained for the case of  photometric imaging with the SGL discussed in  \cite{Turyshev-Toth:2020-photom}.

\subsubsection{Off-image, caustic intersecting the image boundary}

When the astroid is off-image but the astroid caustic intersects the image boundary, the following expression characterizes the intensity at the telescope's focal plane:
  {}
\begin{eqnarray}
I_{\tt out}({\vec x}_i,{\vec x}_0) &=&
 \mu_0\frac{B_{\tt s}}{z^2_0} \Big(\frac{kd^2}{8f}\Big)^2 \frac{ 1}{2\pi \alpha\beta}
  \bigg\{
 \int_{\phi_-}^{\phi_+} \hskip -4pt d\phi''
 \int_{\rho^-_\oplus}^{\rho^+_\oplus}\hskip -4pt d\rho''  {\cal B}_{\tt out}^2({\vec x}_i,{\vec x}'')-\nonumber\\
 &&\hskip 100pt\,-
  \int_{-\phi_1}^{\phi_2} \hskip -4pt d\phi''
 \int_{\rho^-_\oplus}^{\rho_{\tt ac}}\hskip -4pt d\rho'' \Big({\cal B}_{\tt out}^2({\vec x}_i,{\vec x}'')-\Big(\frac{\alpha\beta\rho''}{4\beta_2}\Big){\cal B}_{\tt in}^2({\vec x}_i,{\vec x}'')\Big)
\bigg\},~~~
  \label{eq:intense-y-qu-ap-yo}
\end{eqnarray}
where again $(\phi_1,\phi_2)\equiv(\phi_1(\vec x_0),\phi_2(\vec x_0))$.
We recognize that the first term is identical to that of (\ref{eq:intense-off-quad}), representing the effect of the monopole caustic. The second term looks very similar to that of the last term in (\ref{eq:intense-out*2}), representing the additional flux received at the monopole-suggested circular off-image area.

To evaluate the power received in the telescope's focal plane, we  follow the logic used to develop (\ref{eq:power0}) and (\ref{eq:power0+}), and derive
  {}
\begin{eqnarray}
P_{\tt fp.off}({\vec x}_0)&=& \epsilon_{\tt ee} \int^{2\pi}_0 \hskip -4pt  d\phi_i \int_0^\infty
\hskip 0pt I_{\tt off}({\vec x}_i,{\vec x}_0) \rho_i d\rho_i=\nonumber\\
&=&
\epsilon_{\tt ee} \mu_0\frac{B_{\tt s}}{z^2_0}\frac{\pi d^2 R_\oplus}{2\alpha\beta}\bigg\{  \beta(\rho_0) -
\frac{1}{2\pi R_\oplus} \int_{-\phi_1}^{\phi_2} \hskip -8pt d\phi''  \Big(
\rho_{\tt ac}-\rho^-_\oplus- \Big(\frac{\alpha\beta}{4\beta_2}\Big)\Big(\rho_{\tt ac}^2-(\rho_\oplus^{-})^2+ {\textstyle\frac{1}{2}}\Big(\frac{\alpha\beta}{4\beta_2}\Big)^2(\rho_{\tt ac}^4-(\rho^{-}_\oplus)^4)\Big)
\bigg\}.~~~~~
  \label{eq:power0+}
\end{eqnarray}

Again, assuming that $\phi_1,\phi_2$ and $\phi_0$ are all small, we can evaluate the last integral by taking its value at $\phi''=0$ for $\rho_\oplus$ and $\phi''=\phi_s$ for $\rho_{\tt ac}$. For that, from (\ref{eq:rho++B}) and (\ref{eq:caust-eps-y2}) we have  the following relevant expressions
{}
\begin{eqnarray}
\rho_\oplus(0)
&=&R_\oplus\Big(\frac{\rho_0}{r_\oplus}-1\Big)
\qquad {\rm and} \qquad
\rho_{\tt ac}(0)=\Big(\frac{4\beta_2}{\alpha\beta}\Big)q(\phi_0),
  \label{eq:at-zero2}
\end{eqnarray}
that yield
{}
\begin{eqnarray}
&&
-\frac{1}{2\pi R_\oplus} \int_{-\phi_1}^{\phi_2} \hskip -8pt d\phi''  \Big\{\Big(
\rho_{\tt ac}-\rho^-_\oplus- \Big(\frac{\alpha\beta}{4\beta_2}\Big)\Big(\rho_{\tt ac}^2-(\rho_\oplus^{-})^2+ {\textstyle\frac{1}{2}}\Big(\frac{\alpha\beta}{4\beta_2}\Big)^2(\rho_{\tt ac}^4-(\rho_\oplus^{-})^4)\Big)
\Big\}=\nonumber\\
&=&-\frac{\phi_1+\phi_2}{2\pi}\bigg\{\Big(\frac{\alpha r_\oplus}{4\beta_2}\Big)\Big(\frac{\rho_0}{r_\oplus}-1\Big)^2\Big(1+{\textstyle\frac{1}{2}}\Big(\frac{\alpha r_\oplus}{4\beta_2}\Big)^2\Big(\frac{\rho_0}{r_\oplus}-1\Big)^2\Big)-\Big(\frac{\rho_0}{r_\oplus}-1\Big)+
\Big(\frac{4\beta_2}{\alpha r_\oplus}\Big)\Big(q(\phi_0)-q^2(\phi_0) -{\textstyle\frac{1}{2}}q^4(\phi_0)\Big)
\bigg\}\leq \nonumber\\
&& \hskip 20pt \leq\,
\frac{\phi_1+\phi_2}{2\pi}\bigg\{
\frac{\delta\rho_0}{r_\oplus}+
{\textstyle\frac{5}{6}}\Big(\frac{4\beta_2}{\alpha r_\oplus}\Big)+{\cal O}\Big(\Big(\frac{\delta\rho_0}{r_\oplus}\Big)^2\Big)\bigg\},~~~~~
  \label{eq:power-int}
\end{eqnarray}
where again $\delta \rho_0=r_\oplus-\rho_0$.
As $\rho_0>r_\oplus$, the expression (\ref{eq:power-int}) is negative. This term represents a small increase in the signal in the area outside the image compared to the case of the monopole SGL.

\subsection{Special case of a very large caustic}
\label{sec:large-caust}

In the previous section we have seen that the results depend on  the size of the caustic relative to that of the image.  This is why we consider two special cases: first, when the telescope is still within the image, but the caustic is larger than the image and second, when the telescope is outside the image, but the image is still within the caustic boundary.

\subsubsection{Large caustic with the telescope inside the image}
\label{sec:large-caust-in}

In the case of a very large caustic, $4\beta_2/\alpha \geq \beta R_\oplus=r_\oplus$ (i.e., when $ {\alpha r_\oplus}/{4\beta_2}\leq 1$), when the telescope is inside the image, the intensity of light received in the focal plane of the telescope is given as
  {}
\begin{eqnarray}
I^{\tt in}_{\tt lc}({\vec x}_i,{\vec x}_0) &=&
  \mu_0\frac{B_{\tt s}}{z^2_0} \Big(\frac{kd^2}{8f}\Big)^2   \frac{1}{8\pi\beta_2} \int_0^{2\pi} \hskip -8pt d\phi'' \int_{0}^{\rho_\oplus}\hskip -4pt \rho''d\rho''  {\cal B}_{\tt in}^2({\vec x}_i,{\vec x}'').~~~
  \label{eq:intense-large-c-in}
\end{eqnarray}
Using this result, we compute the power
  {}
\begin{eqnarray}
P_{\tt fp.in.lc}({\vec x}_0)&=& \epsilon_{\tt ee} \int^{2\pi}_0 \hskip -4pt  d\phi_i \int_0^\infty
\hskip 0pt I^{\tt in}_{\tt lc}({\vec x}_i,{\vec x}_0) \rho_i d\rho_i=\nonumber\\
&=&
\epsilon_{\tt ee}\mu_0\frac{B_{\tt s}}{z^2_0} \frac{\pi d^2R_\oplus^2}{8\beta_2} \Big\{1+ {\textstyle\frac{1}{2}}\Big(\frac{\alpha r_\oplus}{4\beta_2}\Big)^2\Big(1+2\Big(\frac{\rho_0}{r_\oplus}\Big)^2\Big)\Big\}\simeq \epsilon_{\tt ee}\mu_0\frac{B_{\tt s}}{z^2_0} \frac{\pi d^2R_\oplus^2}{8\beta_2} \Big\{1+ {\cal O}\Big(\frac{\alpha r_\oplus}{4\beta_2}\Big)^2\Big)\Big\},
  \label{eq:pow-large-c-in}
\end{eqnarray}
where $\rho_0\leq r_\oplus$.
We note that the amount of power received from a source in the case of the extended SGL with the large size of the astroid caustic that is comparable to the size of the image of the source is smaller compared to the case of a monopole PSF by a factor of
  {}
\begin{eqnarray}
P_{\tt fp.in.lc}({\vec x}_0)/P^{\tt mono}_{\tt fp.in.lc}({\vec x}_0)&=&
 \epsilon_{\tt ee}\mu_0\frac{B_{\tt s}}{z^2_0} \frac{\pi d^2R_\oplus^2}{8\beta_2} \Bigg/ \epsilon_{\tt ee} \mu_0\frac{B_{\tt s}}{z^2_0}\frac{\pi d^2 R_\oplus}{2\alpha\beta}=
 \frac{\alpha r_\oplus}{4\beta_2}\leq 1.~~~
  \label{eq:pow-factor}
\end{eqnarray}
Again, from (\ref{eq:4beta2}), we estimate
\begin{eqnarray}
\frac{\alpha r_\oplus}{4\beta_2}&=&\frac{({\overline z}/z_0)R_\oplus}{2J_2 \Big(\frac{R^2_\odot }{\sqrt{2r_g\bar z}}\Big)\sin^2\beta_s}=\frac{2.33}{\sin^2\beta_s}\Big(\frac{\overline z}{650~{\rm AU}}\Big)^\frac{3}{2}\Big(\frac{30~{\rm pc}}{z_0}\Big)\Big(\frac{R_\oplus}{6378\,{\rm km}}\Big).
\label{eq:4beta22*}
\end{eqnarray}

\subsubsection{Large caustic with the telescope outside the image}
\label{sec:large-caust-out}

In the case of a very large caustic, $4\beta_2/\alpha \geq \beta R_\oplus=r_\oplus$, but with the telescope is outsde the image, the intensity of light received at the focal plane of the imaging telescope is given as
  {}
\begin{eqnarray}
I^{\tt out}_{\tt lc}({\vec x}_i,{\vec x}_0) &=&
 \mu_0\frac{B_{\tt s}}{z^2_0} \Big(\frac{kd^2}{8f}\Big)^2   \int_{\phi_-}^{\phi_+} \hskip -8pt d\phi''  \int_{\rho^-_\oplus}^{\rho^+_\oplus}\hskip -4pt  \rho''d\rho''  {\cal A}_{\tt in}^2({\vec x}_i,{\vec x}'')=
  \mu_0\frac{B_{\tt s}}{z^2_0} \Big(\frac{kd^2}{8f}\Big)^2   \frac{1}{8\pi\beta_2} \int_{\phi_-}^{\phi_+} \hskip -8pt d\phi''  \int_{\rho^-_\oplus}^{\rho^+_\oplus}\hskip -4pt  \rho''d\rho''  {\cal B}_{\tt in}^2({\vec x}_i,{\vec x}'').~~~
  \label{eq:intense-large-c-out}
\end{eqnarray}
Using this result, we compute the power
  {}
\begin{eqnarray}
P_{\tt fp.out.lc}({\vec x}_0)&=& \epsilon_{\tt ee} \int^{2\pi}_0 \hskip -4pt  d\phi_i \int_0^\infty
\hskip 0pt I^{\tt out}_{\tt lc}({\vec x}_i,{\vec x}_0) \rho_i d\rho_i=
\epsilon_{\tt ee}\mu_0\frac{B_{\tt s}}{z^2_0} \frac{\pi d^2}{8\beta_2}\frac{1}{2\pi} \int_{\phi_-}^{\phi_+} \hskip -8pt d\phi'' \Big((\rho_\oplus^{+})^2-(\rho_\oplus^{-})^2+ {\textstyle\frac{1}{2}}\Big(\frac{\alpha\beta}{4\beta_2}\Big)^2((\rho_\oplus^{+})^4-)\rho_\oplus^{-})^4)\Big)\simeq 
\nonumber\\
&\simeq&
\epsilon_{\tt ee}\mu_0\frac{B_{\tt s}}{z^2_0} \frac{\pi d^2R_\oplus^2}{2\beta_2}\frac{\rho_0}{r_\oplus}\frac{1}{2\pi} \int_{\phi_-}^{\phi_+} \hskip -8pt d\phi'' \, \cos\phi''\sqrt{1-\Big(\frac{\rho_0}{r_\oplus}\Big)^2\sin^2\phi''},
  \label{eq:pow-large-c-out}
\end{eqnarray}
where $\rho_0\geq r_\oplus$ and where the final simplification is applicable in the case
$({\alpha r_\oplus}/{4\beta_2})<1$.

\section{Evaluating the SGL-amplified signals}
\label{sec:signals}

We consider an Earth-like exoplanet, as viewed from the focal region of the SGL, starting at $\sim$548 AU from the Sun, through a thin-lens telescope. The image of an exoplanet in our galactic neighborhood, at distances up to $\sim$30~pc from the Sun, is projected by the SGL to an image area several kilometers in size ($\sim$1.3~km for an Earth-like exoplanet at 30~pc, observed at 650~AU from the Sun). A telescope in the focal region of the SGL, looking back in the direction of the Sun, sees a faint Einstein ring form around the Sun from light reflected and emitted by the exoplanet.

\subsection{Modeling the spectral signal}

To provide estimates for the anticipated photon fluxes from realistic exoplanetary sources when they are imaged with the SGL, we model the spectral signal using our own Sun and the Earth as representative cases. Following \cite{Turyshev-Toth:2020-extend}, we consider a planet that is identical to our Earth, orbiting, at a distance of 1 AU, a star that is identical to our Sun. The total flux received by such a target is the same as the solar irradiance at the top of Earth's atmosphere. In \cite{Turyshev-Toth:2022-mono-SNR} we developed a realistic model of the spectral brightness $B_s(\lambda)$ of the exoplanet that includes longer wavelengths where we added the planetary thermal emission:
{}
\begin{equation}
B_{\tt s}(\lambda)= \tfrac{2}{3}a
\Big(\frac{R_\odot}{ {\rm AU}}\Big)^2 \frac{2hc^2}{\lambda^5\big(e^{hc/\lambda k_B T_\odot}-1\big)}+ \frac{2hc^2}{\lambda^5\big(e^{hc/\lambda k_B T_\oplus}-1\big)},
\label{eq:model-L0IR*}
\end{equation}
where $a$ is the Earth's visible light albedo, $T_\odot=5772$\,K is the temperature of the Sun and $T_\oplus=252$\,K is the effective radiating temperature of the Earth.

Expression (\ref{eq:intense-y-qu-ap-y}) allows us to compute the power received from the resolved source.
For an actual astrophysical source, $B_s({\vec x}',\lambda)$ is, of course, an arbitrary function of the coordinates ${\vec x}'$ and thus the integral (\ref{eq:pow-blur}) can only be evaluated numerically. However, we can obtain an analytic result in the simple case of a disk of uniform brightness, characterized by ${B}_s({\vec x}',\lambda) ={B}_s(\lambda)$. In this case, we integrate (\ref{eq:intense-y-qu-ap-y}):
  {}
\begin{eqnarray}
I_{\tt }({\vec x}_i,{\vec x}_0,\lambda) &=&
 \mu_0\frac{B_{\tt s}(\lambda)}{z^2_0} \Big(\frac{kd^2}{8f}\Big)^2   \frac{1}{\alpha\beta}\bigg\{  \frac{1}{2\pi}\int_0^{2\pi} \hskip -8pt d\phi'' \int_0^{\rho_\oplus}\hskip -4pt d\rho'' \hat {\cal B}_{\tt out}^2({\vec x}_i,{\vec x}'')- \nonumber\\
 &&\hskip 100pt -\,
\frac{1}{2\pi} \int_0^{2\pi} \hskip -8pt d\phi'' \int_{0}^{\rho_{\tt ac}}\hskip -4pt d\rho''  \Big(
\hat {\cal B}_{\tt out}^2({\vec x}_i,{\vec x}'')- \Big(\frac{\alpha\beta\rho''}{4\beta_2}\Big)\hat{\cal B}_{\tt in}^2({\vec x}_i,{\vec x}'')\Big)\bigg\},~~~
  \label{eq:intense-y-qu-ap-y=}
\end{eqnarray}
where $\hat {\cal B}_{\tt out}({\vec x}_i,{\vec x}'')$ and $\hat {\cal B}_{\tt in}({\vec x}_i,{\vec x}'')$ are the truncated Fourier-transformed amplitudes of the EM field. The first of these two quantities is given as
{}
\begin{eqnarray}
\hat {\cal B}_{\tt out}({\vec x}_i,{\vec x}'') &=&
\Big(\frac{2J_1(\alpha d\, \hat u_0(\phi'',\vec x_i))}{\alpha d\, \hat u_0(\phi'',\vec x_i)}\Big)^2+
 \Big(\frac{2J_1(\alpha d\, \hat u_\pi(\phi'',\vec x_i))}{\alpha d\, \hat u_\pi(\phi'',\vec x_i)}\Big)^2,
  \label{eq:intense1*}
\end{eqnarray}
where
$\hat u_0(\phi'',\vec x_i)$ and $\hat u_\pi(\phi'',\vec x_i)$ introduced by (\ref{eq:upm})  with the help of definitions (\ref{eq:alpha-mu})--(\ref{eq:vec_alpha-eta}) and solutions
(\ref{eq:quad-sol0-y})--(\ref{eq:quad-solpi-y}) are given as
{}
\begin{eqnarray}
\hat u_0(\phi'',\vec x_i)&=&
\Big\{{\textstyle\frac{1}{4}}\Big(1-\frac{\rho_i}{f}\Big(\frac{\overline z}{2r_g}\Big)^\frac{1}{2}\Big)^2+\nonumber\\
&&\hskip 0pt +\,
\frac{\rho_i}{f}\Big(\frac{\overline z}{2r_g}\Big)^\frac{1}{2}\cos^2[{\textstyle\frac{1}{2}}\big(\phi''-{\textstyle\frac{1}{2}}\Big(\frac{4\beta_2}{\alpha\beta\rho''}\Big)\sin[2(\phi''-\phi_s)]+{\textstyle\frac{1}{4}}\Big(\frac{4\beta_2}{\alpha\beta\rho''}\Big)^2\sin[4(\phi''-\phi_s)]-\phi_i\big)] \Big\}^\frac{1}{2},
  \label{eq:mono-u-hat-0+} \\
  \hat u_\pi(\phi'',\vec x_i)&=&
\Big\{{\textstyle\frac{1}{4}}\Big(1-\frac{\rho_i}{f}\Big(\frac{\overline z}{2r_g}\Big)^\frac{1}{2}\Big)^2+
\nonumber\\
&&\hskip 0pt +\,
\frac{\rho_i}{f} \Big(\frac{\overline z}{2r_g}\Big)^\frac{1}{2}\sin^2[{\textstyle\frac{1}{2}}\big(\phi''+{\textstyle\frac{1}{2}}\Big(\frac{4\beta_2}{\alpha\beta\rho''}\Big)\sin[2(\phi''-\phi_s)]+{\textstyle\frac{1}{4}}\Big(\frac{4\beta_2}{\alpha\beta\rho''}\Big)^2\sin[4(\phi''-\phi_s)]-\phi_i\big)]\Big\}^\frac{1}{2}.
  \label{eq:mono-u-hat-pi+}
\end{eqnarray}
Similarly, for $\hat {\cal B}_{\tt in}({\vec x}_i,{\vec x}'') $ we have
{}
\begin{eqnarray}
\hat {\cal B}_{\tt in}({\vec x}_i,{\vec x}'') &=&
\Big(\frac{2J_1(\alpha d\, \hat u^{\tt in}_0(\phi'',\vec x_i))}{\alpha d\, \hat u^{\tt in}_0(\phi'',\vec x_i)}\Big)^2+
 \Big(\frac{2J_1(\alpha d\, \hat u^{\tt in}_\frac{\pi}{2}(\phi'',\vec x_i))}{\alpha d\, \hat u^{\tt in}_\frac{\pi}{2}(\phi'',\vec x_i)}\Big)^2+
 \nonumber\\
&&\hskip 20pt +\,
\Big(\frac{2J_1(\alpha d\, \hat u^{\tt in}_\pi(\phi'',\vec x_i))}{\alpha d\, \hat u^{\tt in}_\pi(\phi'',\vec x_i)}\Big)^2+
 \Big(\frac{2J_1(\alpha d\, \hat u^{\tt in}_\frac{3\pi}{2}(\phi'',\vec x_i))}{\alpha d\, \hat u^{\tt in}_\frac{3\pi}{2}(\phi'',\vec x_i)}\Big)^2,
  \label{eq:intense12*}
\end{eqnarray}
where
$\hat u^{\tt in}_0(\phi'',\vec x_i), \hat u^{\tt in}_\frac{\pi}{2}(\phi'',\vec x_i), \hat u^{\tt in}_\pi(\phi'',\vec x_i)$ and $\hat u^{\tt in}_\frac{3\pi}{2}(\phi'',\vec x_i)$ introduced by (\ref{eq:upm})  with the help of definitions (\ref{eq:alpha-mu})--(\ref{eq:vec_alpha-eta}) and solutions
(\ref{eq:quad-sol0})--(\ref{eq:quad-sol3pi2}) are given as
{}
\begin{eqnarray}
\hat u^{\tt in}_0(\phi'',\vec x_i)&=&
\Big\{{\textstyle\frac{1}{4}}\Big(1-\frac{\rho_i}{f}\Big(\frac{\overline z}{2r_g}\Big)^\frac{1}{2}\Big)^2+
\frac{\rho_i}{f}\Big(\frac{\overline z}{2r_g}\Big)^\frac{1}{2}\cos^2[{\textstyle\frac{1}{2}}\big(\phi_s+\frac{\alpha\beta\rho''}{4\beta_2}\sin(\phi''-\phi_s)-{\textstyle\frac{1}{2}}\Big(\frac{\alpha\beta\rho''}{4\beta_2}\Big)^2\sin[2(\phi''-\phi_s)]+
\nonumber\\
&&+\,
{\textstyle\frac{1}{24}}\Big(\frac{\alpha\beta\rho''}{4\beta_2}\Big)^3\Big(9\sin[\phi''-\phi_s]+5\sin[3(\phi''-\phi_s)]\Big)-{\textstyle\frac{1}{2}}\Big(\frac{\alpha\beta\rho''}{4\beta_2}\Big)^4\sin[2(\phi''-\phi_s)]-\phi_i\big)] \Big\}^\frac{1}{2},
  \label{eq:mono-u-hat-0+in} \\
  \hat u^{\tt in}_\frac{\pi}{2}(\phi'',\vec x_i)&=&
\Big\{{\textstyle\frac{1}{4}}\Big(1-\frac{\rho_i}{f}\Big(\frac{\overline z}{2r_g}\Big)^\frac{1}{2}\Big)^2+
\frac{\rho_i}{f}\Big(\frac{\overline z}{2r_g}\Big)^\frac{1}{2}\cos^2[{\textstyle\frac{1}{2}}\big({\textstyle\frac{\pi}{2}}+ \phi_s+\frac{\alpha\beta\rho''}{4\beta_2}\cos(\phi''-\phi_s)+{\textstyle\frac{1}{2}}\Big(\frac{\alpha\beta\rho''}{4\beta_2}\Big)^2\sin[2(\phi''-\phi_s)]+\nonumber\\
&&+\,
{\textstyle\frac{1}{24}}\Big(\frac{\alpha\beta\rho''}{4\beta_2}\Big)^3\Big(9\cos[\phi''-\phi_s]-5\cos[3(\phi''-\phi_s)]\Big)+{\textstyle\frac{1}{2}}\Big(\frac{\alpha\beta\rho''}{4\beta_2}\Big)^4\sin[2(\phi''-\phi_s)]-\phi_i\big)] \Big\}^\frac{1}{2},
  \label{eq:mono-u-hat-p2+in} \\
  \hat u^{\tt in}_\pi(\phi'',\vec x_i)&=&
\Big\{{\textstyle\frac{1}{4}}\Big(1-\frac{\rho_i}{f}\Big(\frac{\overline z}{2r_g}\Big)^\frac{1}{2}\Big)^2+
\frac{\rho_i}{f} \Big(\frac{\overline z}{2r_g}\Big)^\frac{1}{2}\sin^2[{\textstyle\frac{1}{2}}\big(\phi_s-\frac{\alpha\beta\rho''}{4\beta_2}\sin(\phi''-\phi_s)-{\textstyle\frac{1}{2}}\Big(\frac{\alpha\beta\rho''}{4\beta_2}\Big)^2\sin[2(\phi''-\phi_s)]-\nonumber\\
&&-\,
{\textstyle\frac{1}{12}}\Big(\frac{\alpha\beta\rho''}{4\beta_2}\Big)^3\sin[\phi''-\phi_s]\Big(7 +5\cos[2(\phi''-\phi_s)]\Big)-{\textstyle\frac{1}{2}}\Big(\frac{\alpha\beta\rho''}{4\beta_2}\Big)^4\sin[2(\phi''-\phi_s)]]-\phi_i\big)]\Big\}^\frac{1}{2},
  \label{eq:mono-u-hat-pi+in}  \\
  \hat u^{\tt in}_\frac{3\pi}{2}(\phi'',\vec x_i)&=&
\Big\{{\textstyle\frac{1}{4}}\Big(1-\frac{\rho_i}{f}\Big(\frac{\overline z}{2r_g}\Big)^\frac{1}{2}\Big)^2+
\frac{\rho_i}{f} \Big(\frac{\overline z}{2r_g}\Big)^\frac{1}{2}\sin^2[{\textstyle\frac{1}{2}}\big({\textstyle\frac{\pi}{2}} +\phi_s-\frac{\alpha\beta\rho''}{4\beta_2}\cos(\phi''-\phi_s)+{\textstyle\frac{1}{2}}\Big(\frac{\alpha\beta\rho''}{4\beta_2}\Big)^2\sin[2(\phi''-\phi_s)]-\nonumber\\
&&-\,
{\textstyle\frac{1}{12}}\Big(\frac{\alpha\beta\rho''}{4\beta_2}\Big)^3\cos[\phi''-\phi_s]\Big(7-5\cos[2(\phi''-\phi_s)]\Big)+{\textstyle\frac{1}{2}}\Big(\frac{\alpha\beta\rho''}{4\beta_2}\Big)^4\sin[2(\phi''-\phi_s)]-\phi_i\big)]\Big\}^\frac{1}{2}.
  \label{eq:mono-u-hat-3pi2+in}
\end{eqnarray}

The photon count density (per unit time, unit wavelength, and unit area) that corresponds to (\ref{eq:intense-y-qu-ap-y=}) can be readily calculated:
\begin{align}
Q_{\tt }({\vec x}_i,{\vec x}_0,\lambda)=\frac{\lambda}{hc}I_{\tt }({\vec x}_i,{\vec x}_0,\lambda).
\label{eq:Qsignal}
\end{align}
This quantity is of primary interest as it forms the basis for calculating stochastic shot noise, which results from the quantized nature of light.

\subsection{Spectral signal from the solar corona}

The Einstein ring that forms around the Sun from light emitted by the distant observational target appears on the bright solar corona background. The corona background may be removable (its brightness may be accurately estimated or measured) but as light is quantized, some non-removable stochastic (Poisson) noise inevitably remains.

We model the spectral corona brightness as
{}
\begin{eqnarray}
B_{\tt cor}(\theta,\lambda)&=&
10^{-12}
\frac{2hc^2}{\lambda^\star\lambda^4\big(e^{hc/\lambda k T_\odot}-1\big)}\Big[3.670 \Big(\frac{\theta_0}{\theta}\Big)^{18}+1.939\Big(\frac{\theta_0}{\theta}\Big)^{7.8}+ 5.51\times 10^{-2} \Big(\frac{\theta_0}{\theta}\Big)^{2.5}\Big] ~~   \frac{\rm W}{\mu{\rm m}\,{\rm m}^2\,{\rm sr}}.
\label{eq:model-th-lam*}
\end{eqnarray}

We evaluate the spectral intensity distribution due to corona light, as seen by an imaging telescope, similarly to (\ref{eq:pow-blur}):
  {}
\begin{eqnarray}
I_{\tt cor}({\vec x}_i,\lambda) =
\Big(\frac{kd^2}{8f}\Big)^2
\int_0^{2\pi}\hskip -4pt d\phi'\int_{\theta_{\tt 0}}^{\infty}\hskip -4pt \theta' d\theta'  \, B_{\tt cor}\big(\theta',\lambda\big) \Big( \frac{2
J_1\big(kd\, \hat u(\vec x',\vec x_i)\big)}{kd\, \hat u(\vec x',\vec x_i)}\Big)^2,
  \label{eq:pow-cor*}
\end{eqnarray}
where $\theta_0=R_\odot/{\overline z}$ corresponds to the solar disk, blocked by a coronagraph. Following \cite{Turyshev-Toth:2019-extend}, we introduce the corona spatial frequency, $\vec \alpha_c$ as
{}
\begin{eqnarray}
\vec \alpha_c=k\frac{\vec x'}{\overline z}\equiv k\frac{\rho'}{\overline z}\,\vec n'=k\theta'\,\vec n'.
  \label{eq:upmAc}
\end{eqnarray}
Using (\ref{eq:alpha-mu}) to represent $\vec \eta_i=\eta_i\,\vec n_i$, we offer a wavelength-independent definition for the corona spatial frequency in the form
{}
\begin{eqnarray}
\hat u(\vec x',\vec x_i)=|\vec \alpha_c+\vec \eta_i|/2k=
 \Big\{{\textstyle\frac{1}{4}}\Big(\theta' -\frac{\rho_i}{f}\Big)^2+\theta'\frac{\rho_i}{f}\cos^2[{\textstyle\frac{1}{2}}(\phi'-\phi_i)]\Big\}^\frac{1}{2}.
  \label{eq:upmA}
\end{eqnarray}

The photon count density per unit time, unit wavelength, unit area, according to (\ref{eq:pow-cor*}) is
\begin{align}
Q_{\tt cor}({\vec x}_i,\lambda)=\frac{\lambda}{hc}I_{\tt cor}({\vec x}_i,\lambda).
\label{eq:Qcorona}
\end{align}

Integrating this result over the entire Einstein ring yields the corona's spectral density. 

\subsection{Sensitivity at optical and near-IR wavelengths}
\label{sec:SNR}

The magnitude of non-removable stochastic shot noise is proportional to the square root of the corresponding photon count density. The SGL SNR can be obtained by taking the ratio of the signal (photon count from the Einstein ring) to the square root of the total photon count:
\begin{align}
{\tt SNR}(\vec{x}_0,\vec{x}_i,\lambda)\simeq
\frac{Q(\vec{x}_0,\vec{x}_i,\lambda)}{\sqrt{Q_{\tt cor}(\vec{x}_0,\vec{x}_i,\lambda)}},
\label{eq:theSNR}
\end{align}
where we used the approximation $Q_{\tt cor}\gg Q$, reflecting a faint Einstein ring on a bright corona background.

To compute the SNR for a given integration time $t$, sensor (pixel) area $A$ and spectral channel bandwidth $\lambda$, the quantities in the numerator and under the square root in the denominator must be integrated over these quantities:
{}
\begin{align}
{\tt SNR}\big({\vec x}_0,\Delta t, \Delta \lambda,A_{\tt }\big)= \ddfrac{ \int_{t_1}^{t_2} dt \int_{\lambda_1}^{\lambda_2} d\lambda \iint_A d^2{\vec x}_i~Q(\vec{x}_0,\vec{x}_i,\lambda)}{\Big[\int_{t_1}^{t_2} dt \int_{\lambda_1}^{\lambda_2} d\lambda \iint_A d^2{\vec x}_i~Q_{\tt cor}(\vec{x}_0,\vec{x}_i,\lambda)\Big]^\frac{1}{2}}.
\end{align}

This integral can be replaced by a simple multiplication, provided that the photon flux is constant over the integration time $\Delta t$, the sensor pixel size $A_{\tt pix}$, centered on $\vec{x}_i$, is small, and the spectral channel bandwidth $\Delta\lambda$ centered on $\lambda$ is narrow:
\begin{align}
{\tt SNR}\big({\vec x}_0,\Delta t, \Delta \lambda,A_{\tt }\big)={\tt SNR}(\vec x_0,\vec x_{i}, t,\lambda)\sqrt{A_{\tt pix}~\Delta\lambda~\Delta t}.
\end{align}

\subsection{The deconvolution penalty and simulation results}

The SGL projects a blurred image into the image plane. A sharp image can be restored by unscrambling the blurred signal through the process of deconvolution. This deconvolution inevitably amplifies noise.

To see why this is the case, consider that both the blurring (convolution) and deconvolution are linear processes that map $N$ source pixels $O_i$ into $N$ image pixels $I_j$ or vice versa:
\begin{align}
I_j&{}=\sum_{i=1}^NC_{ij}O_i
\qquad \Rightarrow \qquad
O_i{}=\sum_{j=1}^NC_{ij}^{-1}I_j,
\end{align}
where the matrix $C_{ij}$ is the convolution matrix and it is determined by the PSF:
\begin{align}
C_{ij}={\tt PSF}(\vec{x}_j,\vec{x}'_i).
\end{align}

Without even knowing the specific form of $C_{ij}$, we can quickly draw qualitative conclusions. A blur-free image would be produced by $C_{ij}=\delta_{ij}$, i.e., the identity matrix. Blur mixes up the signal but does not add or remove light. This implies that the actual $C_{ij}$ will have diagonal elements that are reduced from unity, while the off-diagonal elements will increase from 0, but remain small. In short, at any given pixel in the image plane, less light is received from the ``directly imaged'' pixel, while stray light is received from the other pixels of the source.
Deconvolution removes this straight light, so it is essentially a subtractive process. However, the nature of stochastic noise is such that it is always root-square-added. Consequently, as the blur-free signal is restored, noise is amplified by the deconvolution process.

If we could compute an inverse of $C_{ij}$, this conclusion could be explicitly quantified and validated. Unfortunately, given that the PSF is represented by an integral for which at best semianalytical representations are available, direct inversion of the convolution matrix is not practical. However, in \cite{Toth-Turyshev:2020}, we were able to develop a model for the deconvolution matrix that proved useful and reliable for estimating the deconvolution penalty in the case of the monopole PSF. This model was presented in the form
\begin{align}
\tilde{C}_{ij}=\frac{4}{\pi\alpha d}(\mu\delta_{ij}+\nu U_{ij}),
\label{eq:Cij-}
\end{align}
where $U_{ij}$ is the everywhere-one matrix, $\mu\simeq 1$ and $\nu\simeq\ln(\sqrt{2}+1)d/D$, where $D$ is the image size, assuming that the image is sampled at regular intervals in the form of $\sqrt{N}\times \sqrt{N}$ pixels. This matrix is analytically invertible, allowing us to reach the following estimate for the change in the signal-to-noise ratio between the convolved (blurred) signal ${\tt SNR}_{\tt C}$ and the recovered (deconvolved) signal ${\tt SNR}_{\tt R}$ \cite{Turyshev-Toth:2022-mono-SNR}:
\begin{align}
\frac{{\tt SNR}^{\tt mono}_{\tt R}}{{\tt SNR}_{\tt C}}\simeq 0.891\frac{D}{d\sqrt{N}}.
\end{align}

We can extend this estimate to incorporate the deconvolution penalty in the quadrupole case, albeit with a caveat.

In Appendix~\ref{app:Pfp}, we develop an estimate for the ratio of the directly imaged signal in the presence of the quadrupole moment and the corresponding monopole-only signal:
\begin{align}
P^{\tt quad}_{\tt dir}=P^{\tt mono}_{\tt dir}\Big(\frac{\alpha\frac{1}{2} d}{4\beta_2}\Big).
\label{eq:mono-to-quad}
\end{align}
We can incorporate this relationship by scaling the factor $\mu$ in (\ref{eq:Cij-}). This implies scaling the deconvolution penalty:
\begin{align}
\frac{{\tt SNR}^{\tt quad}_{\tt R}}{{\tt SNR}_{\tt C}}\simeq 0.891\Big(\frac{\alpha\frac{1}{2} d}{4\beta_2}\Big)\frac{D}{d\sqrt{N}}.
\label{eq:penalty}
\end{align}

Despite the simplicity of the derivation, this result turns out to be a surprisingly reliable estimator of the actual deconvolution penalty. To investigate this, we numerically evaluated a large number of test cases using an image of the Earth as a stand-in for an exoplanet.

\begin{figure}
\includegraphics{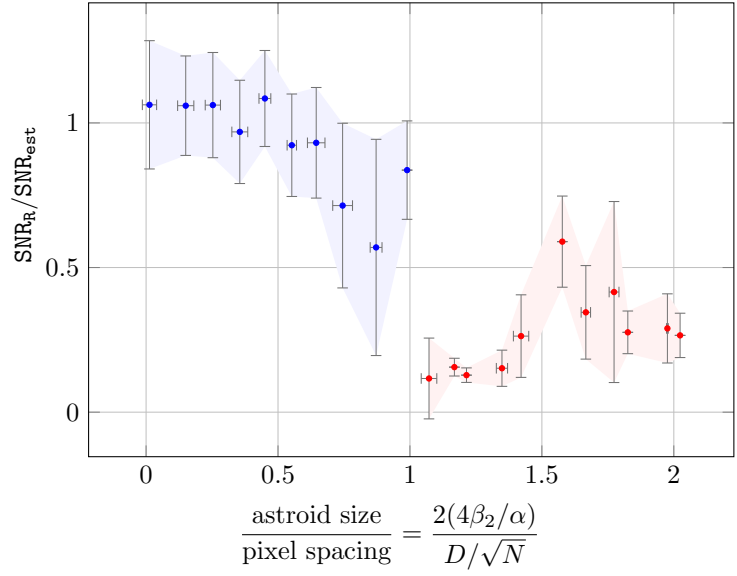}
\caption{\label{fig:paverage}The ratio of the post-deconvolution signal-to-noise ratio ${\tt SNR}_{\tt R}$ to the estimated SNR computed using Eq. (\ref{eq:penalty}). The horizontal axis represents the size of the caustic projection of the quadrupole lens in comparison with the sampling interval, or pixel spacing, in the image plane. The leftmost data point corresponds to the monopole lens, with the caustic increasing in size towards the right. The value of 1 on the horizontal axis represents the case when the size of the caustic, defined as the distance between diagonal cusps, becomes as large as the spacing between the centers of adjacent image pixels.}
\end{figure}
\
Specifically, we evaluated a total number of 5,940 simulation cases, parameterized as follows: $z_0\in\{1.5~{\rm pc}, 10~{\rm pc}, 30~{\rm pc}\}$; ${\overline z}\in\{650~{\rm AU}, 900~{\rm AU}, 1200~{\rm AU}\}$; $d\in\{1~{\rm m}, 2~{\rm m}, 3~{\rm m}, 5~{\rm m}, 10~{\rm m}\}$; $\sqrt{N}\in\{128,256,512,1024\}$; $0\le\sin\beta_s\le 1$ in intervals of $0.1$; and total integration times of 1, 2 and 5 years using a realistic solar corona model at $\lambda=1~\mu$m.
Each simulation began with a monochrome image of the Earth, which was convolved with the quadrupole SGL PSF using Fourier convolution. Gaussian noise corresponding to the solar corona shot noise was added to the convolved image. The image was then deconvolved (again using Fourier deconvolution) and the deconvolved image was compared against the original, to numerically assess the SNR.

The results are depicted in Fig.~\ref{fig:paverage}. As explained in the figure caption, the vertical axis represents the ratio of observed (in simulation) to calculated SNR, whereas the horizontal axis represents the ratio of caustic size to pixel spacing. The plot is based on 1756 simulation cases. These were selected using three criteria:
\begin{inparaenum}[i)]
\item The post-deconvolution SNR had to be greater than 0.5;
\item The estimate (\ref{eq:mono-to-quad}) for the additional deconvolution penalty due to the quadrupole had to be less than 1;
\item The caustic size was to be less than twice the pixel spacing.
\end{inparaenum}
The simulation results were binned for computing averages and standard deviations, as shown. We note that the results, as depicted, are not particularly sensitive to these selection criteria, so the figure is truly representative.

In all cases, we found that so long as the pixel-to-pixel spacing, $D/\sqrt{N}$, was greater than the cusp-to-cusp distance of the quadrupole caustic, $2(4\beta_2/\alpha)$, the observed deconvolution penalty was well modeled by (\ref{eq:penalty}). The estimated penalty remained almost always within a factor-of-two of the observed penalty, but usually it was much closer to the observed value. In fact, for the left-hand side of Fig.~\ref{fig:paverage}, overall average is $0.98\pm 0.25$, indicating that Eq.~(\ref{eq:penalty}) is a robust predictor of the SNR of an image recovered through deconvolution.

When the cusp-to-cusp distance approached or exceeded the pixel spacing, however, the deconvolution penalty rapidly and dramatically increased. This can be understood when we consider the nature of the astroid caustic: the caustic boundary, the cusps in particular, are very bright. When pixels are packed too tightly, this means that much of the light from any source pixels is now deposited outside the directly imaged pixel in the image plane. This violates the conditions for our naive estimate of the deconvolution penalty. Indeed, in these cases, the predictor fared poorly: for the right-hand side of Fig.~\ref{fig:paverage}, the overall average is $0.25\pm 0.20$, indicating that the simulation results yielded SNRs that were significantly worse than predicted, and varied unpredictably in magnitude.

These results indicate that practical imaging with the SGL in the presence of the quadrupole moment is subject to the following two conditions:
\begin{align}
\frac{D}{\sqrt{N}}&{}>2\frac{4\beta_2}{\alpha}, \label{eq:maxres}\\
{\tt SNR}_{\tt C}&{}>1.122\Big(\frac{4\beta_2}{\alpha\frac{1}{2} d}\Big) \frac{d\sqrt{N}}{D} {{\tt SNR}^{\tt quad}_{\tt R}},\label{eq:minSNR}
\end{align}
where ${{\tt SNR}^{\tt quad}_{\tt R}}$ is now the desired post-deconvolution SNR.

The first of these conditions, (\ref{eq:maxres}), determines the maximum image resolution that can be reliably achieved before the deconvolution results degrade drastically.

In contrast, the second condition, (\ref{eq:minSNR}), informs on the minimum required pre-deconvolution SNR needed to achieve a desired post-deconvolution result. This, in turn, can be used to estimate the required integration times.

As an example, let us consider a potential Earth-like target in a solar system situated 1.5~pc from ours, in a direction that is $\sim 17.5^\circ$ from the solar polar direction, such that $\sin\beta_s=0.3$. The corresponding caustic size is $2(4\beta_2/\alpha)\sim 25.9$ meters, whereas the projected image of the Earth-like exoplanet is $D\sim 26.8$~km in diameter. This suggests that in this case, a megapixel scale image ($\sqrt{N}\lesssim 1035$) may be achievable. However, the required integration time may be excessive and indeed, simulation results show that this case is very marginal. On the other hand, an $512\times 512$ image can be reliably obtained, using an integration time of just over 4 years with a single 1-meter telescope. Simulation results confirm this although in this particular case, suggesting a somewhat longer ($\gtrsim 5$ years) cumulative integration time.

Finally, we should mention that when we look at cases of excessively large caustics (e.g., when the observation target is near the solar equatorial plane, so that the contribution of the quadrupole moment is the largest), behind the unpredictable behavior there appear to be possible patterns. Occasionally, simulation results were unexpectedly good, with a much better SNR than we would have expected under the circumstances. Our suspicion is that this is a result of ``aliasing'', a resonance between the selected pixel spacing and the caustic size. We performed some limited experimentation by varying simulation parameters, and while we achieved no definite conclusion, the results seem to support our conjecture. This will need to be investigated in detail in the future, as such aliasing may lead to carefully orchestrated observational strategies that could yield ``super-resolution'' images, images of much better quality than we might otherwise expect, even in the presence of large caustics, obtained using realistic integration times.

\section{Conclusion}
\label{sec:conc}

We have studied the optical properties of the quadrupole PSF of the SGL. We developed an approximate solution for the diffraction integral in two distinct regions: the regions inside and outside the caustic boundary. These solutions were used to study the intensity distribution for light received in the focal plane of an imaging telescope.

Compared to the case when the Sun is treated as a spherically symmetric gravitating body characterized by the monopole axially-symmetric PSF,  the solar quadrupole introduces an astroid caustic. The presence of that caustic does not significantly affect the total power deposited in the telescope's focal plane. In fact, in the majority of practical cases, the total power is reduced by $\lesssim 5\%$. However, the light received in the focal plane is scrambled, with photons scattered across image pixels. This contributes significant blurring above and beyond the blur that is already present due to the spherical aberration that characterizes the monopole lens.

As we are in possession of a PSF that describes the quadrupole lens accurately, a sharp image can be recovered in principle by the process of deconvolution. This, however, amplifies noise. Such noise is inevitably present in the case of an image produced by the SGL, as such an image is formed by light that appears in the form of a faint Einstein ring on top of a very bright solar corona background. Due to the quantized nature of light, even if the corona is accurately modeled and removed, stochastic noise (shot noise) remains.

To estimate the impact of this noise, we used a simplistic model that we developed originally for the monopole PSF. This model captures the qualitative properties of the convolution matrix and allowed us to estimate the ``deconvolution penalty'' accurately. We found that this model can be readily extended to the quadrupole case with an additional scaling of signal levels. Although more work is needed, the results are already encouraging. Deconvolution remains practical so long as the astroid caustic remains smaller than the spacing between pixels, and in this regime, the estimated deconvolution penalty is well matched by numerical simulation results.

Using these results, we were able to express a simple set of conditions that must be satisfied for practical image recovery, using the SGL as a telescope. This leaves only one major task that needs to be accomplished in order to fully model exoplanetary imaging using the SGL: modeling of temporal behavior, including planetary diurnal rotation, its orbital motion and phases of illumination. This work is ongoing; results, when available, will be reported elsewhere.

\begin{acknowledgments}
This work in part was performed at the Jet Propulsion Laboratory, California Institute of Technology, under a contract with the National Aeronautics and Space Administration.
VTT acknowledges the generous support of Plamen Vasilev and other Patreon patrons.
\end{acknowledgments}


\appendix

\section{EM field amplitude inside the caustic region}
\label{app:EMamp}

We present ${\cal B}_{\tt in}^2({\vec x}_i,{\vec x}'')$, which is referenced in Eq.~(\ref{eq:psf-quad_in-AB}), to terms up to ${\cal O}(x^5)$:
{}
 \begin{eqnarray}
{\cal B}_{\tt in}^2({\vec x}_i,{\vec x}'')&=&
\Big( {\cal A}^2_0+ {\cal A}^2_{\pi/2}+ {\cal A}^2_\pi+ {\cal A}^2_{3\pi/2}\Big)\Big(1+ x^2 +
 {\textstyle\frac{1}{4}}\big(9-5\cos[4(\phi''-\phi_s)]\big)\,x^4\Big)+\nonumber\\[2pt]
 &&\hskip -80pt+\,
\Big( {\cal A}^2_{\pi}-{\cal A}^2_{0}\Big)\cos(\phi''-\phi_s)\, x\Big(1 +
 \big(1+{\textstyle\frac{7}{2}}\sin^2(\phi''-\phi_s)\big)\,x^2\Big)+\Big( {\cal A}^2_{\pi/2}-{\cal A}^2_{3\pi/2}\Big)\sin(\phi''-\phi_s)\, x\Big(1 +
 \big(1+{\textstyle\frac{7}{2}}\cos^2(\phi''-\phi_s)\big)\,x^2\Big)+\nonumber\\[2pt]
 &&\hskip -80pt+\,\sin2\beta_2\bigg\{
\Big( {\cal A}_{0}+{\cal A}_{\pi}\Big)\Big( {\cal A}_{\pi/2}+{\cal A}_{3\pi/2}\Big)\Big(481-267\cos[4(\phi''-\phi_s)]\Big)
+\Big( {\cal A}_{0}-{\cal A}_{\pi}\Big)\Big( {\cal A}_{\pi/2}-{\cal A}_{3\pi/2}\Big)148\sin[2(\phi''-\phi_s)]\bigg\} {\textstyle\frac{1}{128}}x^4+\nonumber\\[2pt]
&&\hskip-80pt +\,
2{\cal A}_0{\cal A}_{\pi} \bigg\{\cos\big[8\beta_2 x \cos(\phi''-\phi_s)\big]\Big(1+{\textstyle\frac{1}{2}}x^2\big(1+\sin^2(\phi''-\phi_s)\big)+{\textstyle\frac{1}{64}} x^4\big(97-20\cos[2(\phi''-\phi_s)]-53\cos[4(\phi''-\phi_s)]\big)\Big)+\nonumber\\
&&+\, \sin\big[8\beta_2 x \cos(\phi''-\phi_s)\big]\sin\Big[4\beta_2 x^3 \cos(\phi''-\phi_s)\sin^2(\phi''-\phi_s)\Big)\bigg\}+
\nonumber\\[2pt]
&&\hskip-80pt +\,
2{\cal A}_{\pi/2}{\cal A}_{3\pi/2}\bigg\{ \cos\big[8\beta_2 x \sin(\phi''-\phi_s)\big] \Big(1+{\textstyle\frac{1}{2}}x^2\big(1+\cos^2(\phi''-\phi_s)\big)+{\textstyle\frac{1}{64}} x^4\Big(97+20\cos[2(\phi''-\phi_s)]-53\cos[4(\phi''-\phi_s)]\Big)\Big)+
\nonumber\\
&&+\, \sin\big[8\beta_2 x \sin(\phi''-\phi_s)\big]\sin\Big[4\beta_2 x^3 \sin(\phi''-\phi_s)\cos^2(\phi''-\phi_s)\Big)\bigg\}+
\nonumber\\[2pt]
&&\hskip-80pt +\,
\Big({\cal A}_0{\cal A}_{3\pi/2}+{\cal A}_{\pi/2}{\cal A}_\pi\Big)\bigg\{2\sin\Big[2\beta_2\Big(1+x^2+{\textstyle\frac{1}{4}}x^4\big(1-\cos[4(\phi''-\phi_s)]\big)\Big)\Big]\times
\nonumber\\[2pt]
&&\hskip-00pt\times\,
\cos\Big[4\beta_2x\Big(\cos(\phi''-\phi_s)+\sin(\phi''-\phi_s)\Big)\Big(1-{\textstyle\frac{1}{4}}x^2\sin[2(\phi''-\phi_s)]\Big)\Big]\Big(1+{\textstyle\frac{1}{8}}x^2\big(7+\sin[2(\phi''-\phi_s)]\big)\Big)
-\nonumber\\
&&\hskip-40pt -\,
2\cos\Big[2\beta_2\Big(1+x^2\Big)\Big]
\sin\Big[4\beta_2x\Big(\cos(\phi''-\phi_s)+\sin(\phi''-\phi_s)\Big)\Big(1-{\textstyle\frac{1}{4}}x^2\sin[2(\phi''-\phi_s)]\Big)\Big]\times
\nonumber\\
&&\hskip-0pt\times\,
{\textstyle\frac{1}{2}}x\Big(\cos(\phi''-\phi_s)+\sin(\phi''-\phi_s)\Big)\Big(1+{\textstyle\frac{1}{8}}x^2\big(9+13\sin[2(\phi''-\phi_s)]\big)\Big)\bigg\}+\nonumber\\
&&\hskip-80pt+\,
\Big({\cal A}_0{\cal A}_{3\pi/2}-{\cal A}_{\pi/2}{\cal A}_\pi\Big)\bigg\{2\cos\Big[2\beta_2\Big(1+x^2\Big)\Big]
\sin\Big[4\beta_2x\Big(\cos(\phi''-\phi_s)+\sin(\phi''-\phi_s)\Big)\Big(1-{\textstyle\frac{1}{4}}x^2\sin[2(\phi''-\phi_s)]\Big)\Big]\times
\nonumber\\
&&\hskip-0pt\times\,
\Big(1+{\textstyle\frac{1}{8}}x^2\big(7+\sin[2(\phi''-\phi_s)]\big)\Big)-\nonumber\\
&&\hskip-40pt-\,
2\sin\Big[2\beta_2\Big(1+x^2\Big)\Big]
\cos\Big[4\beta_2x\Big(\cos(\phi''-\phi_s)+\sin(\phi''-\phi_s)\Big)\Big]\times
\nonumber\\
&&\hskip-00pt\times\,
{\textstyle\frac{1}{2}}x\Big(\cos(\phi''-\phi_s)+\sin(\phi''-\phi_s)\Big)\Big(1+{\textstyle\frac{1}{8}}x^2\big(9+13\sin[2(\phi''-\phi_s)]\big)\Big)\bigg\}+\nonumber\\[2pt]
&&\hskip-80pt+\,
\Big({\cal A}_0{\cal A}_{\pi/2}+{\cal A}_\pi{\cal A}_{3\pi/2}\Big)\bigg\{2\sin\Big[2\beta_2\Big(1+x^2+{\textstyle\frac{1}{4}}x^4\big(1-\cos[4(\phi''-\phi_s)]\big)\Big)\Big]\times
\nonumber\\
&&\hskip-00pt\times\,
\cos\Big[4\beta_2x\Big(\cos(\phi''-\phi_s)-\sin(\phi''-\phi_s)\Big)\Big(1+{\textstyle\frac{1}{4}}x^2\sin[2(\phi''-\phi_s)]\Big)\Big]
\Big(1+{\textstyle\frac{1}{8}}x^2\big(7-\sin[2(\phi''-\phi_s)]\big)\Big)-\nonumber\\
&&\hskip-40pt -\,
2\cos\Big[2\beta_2\Big(1+x^2\Big)\Big]
\sin\Big[4\beta_2x\Big(\cos(\phi''-\phi_s)-\sin(\phi''-\phi_s)\Big)\Big(1+{\textstyle\frac{1}{4}}x^2\sin[2(\phi''-\phi_s)]\Big)\Big]\times
\nonumber\\
&&\hskip-0pt\times\,
{\textstyle\frac{1}{2}}x\Big(\cos(\phi''-\phi_s)-\sin(\phi''-\phi_s)\Big)\Big(1+{\textstyle\frac{1}{8}}x^2\big(9-13\sin[2(\phi''-\phi_s)]\big)\Big)\bigg\}+\nonumber\\[2pt]
&&\hskip-80pt+\,
\Big({\cal A}_{\pi}{\cal A}_{3\pi/2}-{\cal A}_0{\cal A}_{\pi/2}\Big)
\bigg\{2\sin\Big[2\beta_2\Big(1+x^2\Big)\Big]
\cos\Big[4\beta_2x\Big(\cos(\phi''-\phi_s)-\sin(\phi''-\phi_s)\Big)\Big]\times
\nonumber\\
&&\hskip-00pt\times\,
{\textstyle\frac{1}{2}}x\Big(\cos(\phi''-\phi_s)-\sin(\phi''-\phi_s)\Big)\Big(1+{\textstyle\frac{1}{8}}x^2\big(9-13\sin[2(\phi''-\phi_s)]\big)\Big)-
\nonumber\\
&&\hskip-40pt-\,
2\cos\Big[2\beta_2\Big(1+x^2\Big)\Big]
\sin\Big[4\beta_2x\Big(\cos(\phi''-\phi_s)-\sin(\phi''-\phi_s)\Big)\Big(1+{\textstyle\frac{1}{4}}x^2\sin[2(\phi''-\phi_s)]\Big)\Big]\times
\nonumber\\
&&\hskip-0pt\times\,
\Big(1+{\textstyle\frac{1}{8}}x^2\big(7-\sin[2(\phi''-\phi_s)]\big)\Big)\bigg\}+
 {\cal O}(x^5),
  \label{eq:psf-quad_in-AB2}
\end{eqnarray}
where $x\equiv \big({\alpha\beta\rho''}/{4\beta_2}\big)<1$ as given by (\ref{eq:quad-small-par}).

\section{The signal power received by an imaging telescope}
\label{app:Pfp}

Assuming that the size of the caustic is smaller than the image size, $4\beta_2/\alpha < r_\oplus$, and the observing telescope positioned at such a distance from the caustic boundary that the caustic is not intersecting the boundary, $\rho_0+4\beta_2/\alpha \leq r_\oplus$, we compute the intensity received by the telescope:
{}
\begin{eqnarray}
I_{\tt }({\vec x}_i,{\vec x}_0) &=&
 \mu_0\frac{B_{\tt s}}{z^2_0} \Big(\frac{kd^2}{8f}\Big)^2  \bigg\{ \int_0^{2\pi} \hskip -8pt d\phi'' \int_{0}^{\rho_{\tt ac}}\hskip -4pt \rho''d\rho''  {\cal A}_{\tt in}^2({\vec x}_i,{\vec x}'')+ \int_0^{2\pi} \hskip -8pt d\phi'' \int_{\rho_{\tt ac}}^{\rho_\oplus}\hskip -4pt \rho''d\rho''  {\cal A}_{\tt out}^2({\vec x}_i,{\vec x}'')\bigg\}=\nonumber\\
&=&
\mu_0\frac{B_{\tt s}}{z^2_0} \Big(\frac{kd^2}{8f}\Big)^2  \bigg\{
\frac{1}{8\pi\beta_2} \int_0^{2\pi} \hskip -8pt d\phi'' \int_{0}^{\rho_{\tt ac}}\hskip -4pt \rho''d\rho'' \, {\cal B}_{\tt in}^2({\vec x}_i,{\vec x}'')+
\frac{1}{2\pi\alpha\beta}\int_0^{2\pi} \hskip -8pt d\phi'' \int_{\rho_{\tt ac}}^{\rho_\oplus}\hskip -4pt d\rho''   {\cal B}_{\tt out}^2({\vec x}_i,{\vec x}'')\bigg\}.~~~
  \label{eq:intense-pp*}
\end{eqnarray}

Equation~(\ref{eq:intense-pp*}) allows us to estimate the power received by  an imaging telescope form a position with the image. With the help of expressions (\ref{eq:int-B1})--(\ref{eq:int-B2}), we have (again, with $q(\phi'')$ from (\ref{eq:caust-eps-y2-q}))
{}
\begin{eqnarray}
P_{\tt fp}({\vec x}_0)&=& \epsilon_{\tt ee} \int^{2\pi}_0 \hskip -4pt  d\phi_i \int_0^\infty
\hskip 0pt I_{\tt }({\vec x}_i,{\vec x}_0) \rho_i d\rho_i=\nonumber\\
&=&
\epsilon_{\tt ee} \mu_0\frac{B_{\tt s}}{z^2_0} \frac{\pi d^2}{2} \bigg\{
\frac{2}{8\pi\beta_2} \int_0^{2\pi} \hskip -8pt d\phi''
\int_{0}^{\frac{1}{2}D}\hskip -4pt \rho''d\rho'' \Big\{1+ \Big(\frac{\alpha\beta\rho''}{4\beta_2}\Big)^2+
 {\textstyle\frac{1}{4}}\Big(9-5\cos[4(\phi''+\phi_0-\phi_s)]\Big)\Big(\frac{\alpha\beta\rho''}{4\beta_2}\Big)^4\Big) \Big\}+\nonumber\\
&&\hskip 25pt +\,
\frac{2}{8\pi\beta_2} \int_0^{2\pi} \hskip -8pt d\phi''
\int_{\frac{1}{2}D}^{\rho_{\tt ac}}\hskip -4pt \rho''d\rho'' \Big\{1+ \Big(\frac{\alpha\beta\rho''}{4\beta_2}\Big)^2+
 {\textstyle\frac{1}{4}}\Big(9-5\cos[4(\phi''+\phi_0-\phi_s)]\Big)\Big(\frac{\alpha\beta\rho''}{4\beta_2}\Big)^4\Big) \Big\}+
  \nonumber\\
&&\hskip 80pt +\,
\frac{1}{2\pi\alpha\beta}\int_0^{2\pi} \hskip -8pt d\phi'' \int_{\rho_{\tt ac}}^{\rho_\oplus}\hskip -4pt d\rho''  \bigg\}=\nonumber\\
&=&
\epsilon_{\tt ee} \mu_0\frac{B_{\tt s}}{z^2_0} \frac{\pi d^2}{2} \bigg\{
\frac{1}{4\beta_2}
\Big(\frac{4\beta_2}{\alpha\beta}\Big)^2\frac{1}{2\pi}\int_0^{2\pi} \hskip -8pt d\phi''  \Big\{
q^2(\phi'')+ {\textstyle\frac{1}{2}}q^4(\phi'')+
 {\textstyle\frac{1}{4}}
 \Big(9-5\cos[4(\phi''+\phi_0-\phi_s)]\Big)
q^6(\phi'')\Big\} +
\nonumber\\
&&\hskip 60pt +\,
\frac{R_\oplus}{2\pi\alpha\beta}\int_0^{2\pi} \hskip -8pt d\phi'' \bigg(\sqrt{1-\Big(\frac{\rho_0}{r_\oplus}\Big)^2\sin^2\phi''}-
\Big(\frac{4\beta_2}{\alpha r_\oplus}\Big)
q(\phi'')\bigg) \bigg\}=
\nonumber\\
&=&
\epsilon_{\tt ee} \mu_0\frac{B_{\tt s}}{z^2_0} \frac{\pi d^2}{2} \bigg\{
\frac{1}{4\beta_2}
\Big(\frac{4\beta_2}{\alpha\beta}\Big)^2 \Big\{0.375+0.082+ 0.039\Big\} +
\frac{R_\oplus}{\alpha\beta} \Big(\epsilon(\rho_0)-0.602\Big(\frac{4\beta_2}{\alpha r_\oplus}\Big) \Big) \bigg\}.
\label{eq:pow-con}
\end{eqnarray}

Assuming that $\big({\alpha {\textstyle\frac{1}{2}}d}/{4\beta_2}\big)\ll1$, we simplify the result (\ref{eq:pow-con}) as
{}
\begin{eqnarray}
P_{\tt fp}({\vec x}_0)&=&
\epsilon_{\tt ee}  \mu_0\frac{B_{\tt s}}{z^2_0} \frac{\pi d^2}{2} \frac{1}{4 \beta_2}\frac{d^2}{4\beta^2 } \bigg\{
1 +  \Big(\Big(\frac{4\beta_2}{\alpha\frac{1}{2} d}\Big)^2 0.496-1\Big) +
\Big(\frac{4\beta_2}{\alpha\frac{1}{2} d}\Big) \Big( \Big(\frac{2r_\oplus}{d}\Big) \epsilon(\rho_0)-0.602\Big(\frac{4\beta_2}{\alpha\frac{1}{2} d}\Big) \Big) \bigg\}=\nonumber\\
&=& P^{\tt quad}_{\tt dir}\bigg\{
1 +  \Big(\Big(\frac{4\beta_2}{\alpha\frac{1}{2} d}\Big)^2 0.496-1\Big) +
\Big(\frac{4\beta_2}{\alpha\frac{1}{2} d}\Big) \Big( \Big(\frac{2r_\oplus}{d}\Big) \epsilon(\rho_0)-0.602\Big(\frac{4\beta_2}{\alpha\frac{1}{2} d}\Big)\Big)  \bigg\}=\nonumber\\
&=&
P^{\tt quad}_{\tt dir}\bigg\{
1 +  \Big(163,865-1\Big) +
 \Big( 725,010\,\epsilon(\rho_0)-198,884 \Big) \bigg\}=
 P^{\tt quad}_{\tt dir}\bigg\{
725,010\,\epsilon(\rho_0)-35,019 \bigg\},
  \label{eq:power-pp*}
\end{eqnarray}
with $\epsilon(\rho_0)$ given by (\ref{eq:eps_r0}) and where we have
\begin{eqnarray}
P^{\tt quad}_{\tt dir}=\epsilon_{\tt ee} \mu_0\frac{B_{\tt s}}{z^2_0} \frac{\pi d^2}{2} \frac{1}{4 \beta_2}\frac{d^2}{4\beta^2 } =
P^{\tt mono}_{\tt dir}\Big(\frac{\alpha\frac{1}{2} d}{4\beta_2}\Big), \qquad {\rm where}\qquad
P^{\tt mono}_{\tt dir}=\epsilon_{\tt ee} \mu_0\frac{B_{\tt s}}{z^2_0} \frac{\pi d^3}{4\alpha \beta^2}.
\label{eq:P_dp_m}
\end{eqnarray}

One can see that the monopole contribution is dominant with quadrupole taking $\sim 5\%$ of the power outside the image defined by the monopole PSF.

As a result, using (\ref{eq:power-pp*}) and (\ref{eq:P_dp_m}), we have the following expression to describe the total power received by a telescope at a particular position within the image
{}
\begin{eqnarray}
P_{\tt fp}({\vec x}_0)&=& P^{\tt quad}_{\tt dir}\Big\{
\Big(\frac{4\beta_2}{\alpha\frac{1}{2} d}\Big)^2 0.496+
\Big(\frac{4\beta_2}{\alpha\frac{1}{2} d}\Big)
\Big( \Big(\frac{2r_\oplus}{d}\Big) \epsilon(\rho_0)-0.602\Big(\frac{4\beta_2}{\alpha\frac{1}{2} d}\Big) \Big)  \Big\}=
\nonumber\\ &=&
P^{\tt quad}_{\tt dir}\Big(\frac{4\beta_2}{\alpha\frac{1}{2} d}\Big) \Big\{
\Big(\frac{2r_\oplus}{d}\Big) \epsilon(\rho_0)-0.106\Big(\frac{4\beta_2}{\alpha\frac{1}{2} d}\Big)\Big\}\equiv
P^{\tt mono}_{\tt dir} \Big(\frac{2r_\oplus}{d}\Big) \Big\{ \epsilon(\rho_0)-0.106\Big(\frac{4\beta_2}{\alpha r_\oplus}\Big)\Big\}=
\nonumber\\ &=&
P^{\tt mono}_{\tt fp}({\vec x}_0)\Big\{ \epsilon(\rho_0)-0.106\Big(\frac{4\beta_2}{\alpha r_\oplus}\Big)\Big\}.
  \label{eq:power-pp4}
\end{eqnarray}

This result quantifies the fact that even in the case of the quadrupole PSF, most of the image photons are still received with the image defined by the monopole PSF.

\end{document}